\documentclass[useAMS,usenatbib]{mn2e}

\usepackage{graphicx}	
\usepackage{amsmath}	
\usepackage{amssymb}	
\usepackage{multicol}        
\usepackage{bm}		
\usepackage{pdflscape}	
\usepackage{multirow,multicol} 

\def\N1316{NGC\,1316}
\def\N1404{NGC\,1404}

\def\4U{4U~1735$-$444}

\def\arcsec{\ifmmode '' \else $''$\fi}

\def\arcsecpoint{\ifmmode ''\!. \else $''\!.$\fi}

\def\kms{\ifmmode {\rm km\ s}^{-1} \else km s$^{-1}$\fi}
\def\Msun{\ifmmode {\rm M}_{\odot} \else M$_{\odot}$\fi}
\def\Lsun{\ifmmode {\rm L}_{\odot} \else L$_{\odot}$\fi}
\def\Zsun{\ifmmode {\rm Z}_{\odot} \else Z$_{\odot}$\fi}

\def\ergscm2{ergs\,s$^{-1}$\,cm$^{-2}$}
\def\icm3{{\rm cm}^{-3}}
\def\icm2{{\rm cm}^{-2}}
\def\qo{\ifmmode q_{\rm o} \else $q_{\rm o}$\fi}
\def\Ho{\ifmmode H_{\rm o} \else $H_{\rm o}$\fi}
\def\ho{\ifmmode h_{\rm o} \else $h_{\rm o}$\fi}

\def\vFWHM{\ifmmode v_{\mbox{\tiny FWHM}} \else
            $v_{\mbox{\tiny FWHM}}$\fi}
\def\CCF{\ifmmode F_{\it CCF} \else $F_{\it CCF}$\fi}
\def\ACF{\ifmmode F_{\it ACF} \else $F_{\it ACF}$\fi}
\def\Halpha{\ifmmode {\rm H}\alpha \else H$\alpha$\fi}
\def\Hbeta{\ifmmode {\rm H}\beta \else H$\beta$\fi}
\def\Hgamma{\ifmmode {\rm H}\gamma \else H$\gamma$\fi}
\def\Hdelta{\ifmmode {\rm H}\delta \else H$\delta$\fi}
\def\Lya{\ifmmode {\rm Ly}\alpha \else Ly$\alpha$\fi}
\def\Lyb{\ifmmode {\rm Ly}\beta \else Ly$\beta$\fi}
\def\Lyg{\ifmmode {\rm Ly}\beta \else Ly$\gamma$\fi}

\def\ciii{\ifmmode {\rm C}\,{\sc iii} \else C\,{\sc iii}\fi}
\def\civ{\ifmmode {\rm C}\,{\sc iv} \else C\,{\sc iv}\fi}
\def\cv{\ifmmode {\rm C}\,{\sc v} \else C\,{\sc v}\fi}
\def\cvi{\ifmmode {\rm C}\,{\sc vi} \else C\,{\sc vi}\fi}

\def\o5007{[O\,{\sc iii}]\,$\lambda5007$}

\def\ovii{O\,{\sc vii}}
\def\oviii{O\,{\sc viii}}

\def\neix{Ne\,{\sc ix}}
\def\nex{Ne\,{\sc x}}

\def\mgxii{Mg\,{\sc xii}}

\def\fexxii-iii{Fe\,{\sc xxii-xxiii}}

\title[From ULX to ULS: NGC 55 ULX]{From ultraluminous X-ray sources to ultraluminous supersoft sources: NGC 55 ULX, the missing link}
\author[C. Pinto et al.]{C. Pinto,$^{1}$\thanks{E-mail:
cpinto@ast.cam.ac.uk} W. Alston$^{1}$, R. Soria,$^{2,3}$ M. J. Middleton,$^{4}$ 
       \newauthor D. J. Walton,$^{1}$  A. D. Sutton,$^{5}$ A. C. Fabian,$^{1}$ H. Earnshaw,$^{6}$ 
        \newauthor R. Urquhart,$^{2}$ E. Kara\,$^{7}$ and T. P. Roberts,$^{6}$ \\
$^{1}$ Institute of Astronomy, Madingley Road, CB3 0HA Cambridge, United Kingdom\\
$^{2}$ International Centre for Radio Astronomy Research, Curtin University, GPO Box U1987, Perth, WA 6845, Australia\\
$^{3}$ Sydney Institute for Astronomy, School of Physics A28, The University of Sydney, Sydney, NSW 2006, Australia\\
$^{4}$ Physics \& Astronomy, University of Southampton, Southampton, Hampshire SO17 1BJ, UK\\
$^{5}$ Astrophysics Office, NASA Marshall Space Flight Center, ZP12, Huntsville, AL 35812, USA\\
$^{6}$ Centre for Extragalactic Astronomy, Durham University, Department of Physics, South Road, Durham DH1 3LE, UK\\
$^{7}$ Department of Astronomy, University of Maryland, College Park, MD 20742-2421, USA}
\begin{document}

\date{Accepted 2017 March 13. Received 2017 March 13 ; in original form 2016 December 16}

\pagerange{\pageref{firstpage}--\pageref{lastpage}} \pubyear{2016}

\maketitle

\label{firstpage}

\begin{abstract}
In recent work with high-resolution grating spectrometers (RGS) 
aboard XMM-\textit{Newton} Pinto et al. (2016) 
have discovered that two bright and archetypal ultraluminous X-ray sources (ULXs)
have strong relativistic winds in agreement with theoretical predictions
of high accretion rates.
It has been proposed that such winds can become optically thick 
enough to block and reprocess the disk X-ray photons almost entirely, 
making the source appear as a soft thermal emitter or 
ultraluminous supersoft X-ray source (ULS).
To test this hypothesis we have studied a ULX where the wind 
is strong enough to cause significant absorption of the hard X-ray continuum: NGC 55 ULX. 
The RGS spectrum of NGC 55 ULX shows a wealth of emission and absorption 
lines blueshifted by significant fractions of the light speed 
$(0.01-0.20)c$ indicating the presence of a powerful wind. 
The wind has a complex dynamical structure with the ionization state
increasing with the outflow velocity, which may indicate
launching from different regions of the accretion disk.
The comparison with other ULXs such as NGC 1313 X-1 and NGC 5408 X-1
suggests that NGC 55 ULX is being observed at higher inclination.
The wind partly absorbs the source flux above 1\,keV,
generating a spectral drop similar to that observed in ULSs.
The softening of the spectrum at lower ($\sim$ Eddington) luminosities
and the detection of a soft lag agree with the scenario of wind clumps 
crossing the line of sight, partly absorbing and reprocessing 
the hard X-rays from the innermost region.
\end{abstract}


\section{Introduction}
\label{sec:intro}

Ultraluminous X-ray sources (ULXs) are bright, point-like, off-nucleus,
extragalactic sources with X-ray luminosities above $10^{39}$ erg/s
that result from accretion onto a compact object. 
Previous studies have shown evidence of
accretion onto neutron stars with strong magnetic fields 
(e.g. Bachetti et al. 2014, Israel et al. 2016a,b, Fuerst et al. 2016).
{It is suggested that ULXs are also powered by}
accretion onto stellar-mass black holes $(<100 M_{\odot})$ at or in
excess of the Eddington limit (e.g. King et al. 2001, 
Poutanen et al. 2007, Gladstone et al. 2009, Middleton et al. 2013, Liu et al. 2013)  
or accretion at more sedate Eddington ratios onto intermediate mass black holes
($10^{3-5} M_{\odot}$, e.g. Greene and Ho 2007, Farrell et al. 2009, Webb at al. 2012, Mezcua et al. 2016). 

Ultraluminous supersoft sources (ULSs) are defined by a thermal spectrum 
with colour temperature $\sim$ 0.1 keV, bolometric luminosity $\sim$ a few 
$10^{39}$ erg/s, and almost no emission above 1 keV (Kong \& Di Stefano 2003).
Deep exposures of ULSs have shown the presence of a hard tail,
possibly due to a disk-like emission similar to ULXs 
(see, e.g, Urquhart and Soria 2016).
Classical X-ray binaries and ULXs,  
either have broad-band emission over the 1--10 keV range or
a peak disk-blackbody temperature {of a few keV}.
Alternative models have tried to describe the X-ray 
spectra of ULSs with accreting {intermediate-mass and} stellar-mass black holes 
like for ULXs or with extreme supersoft sources
powered by surface-nuclear-burning on white dwarf accretors 
(see, e.g., Di Stefano \& Kong 2004, Soria \& Kong 2016, Feng et al. 2016, 
and references therein).

ULXs and ULSs were initially considered different physical types of systems. 
However, on the one hand, we have seen strong evidence of winds in {classical} ULXs
(Middleton et al. 2015b, Pinto et al. 2016). 
On the other hand, some ULSs (e.g. in M\,101 and NGC\,247) 
exhibit a harder (fast variable) X-ray tail (e.g., Urquhart and Soria 2016), 
suggesting that some of the X-ray photons occasionally get 
through a rapidly changing wind 
(e.g., Middleton et al. 2011 and Takeuchi et al. 2013). 
{It is therefore speculated that} ULXs and ULSs are simply 
two types of super-Eddington accretors, 
with respectively geometrically thinner and thicker outflows
along the line of sight due to different viewing angles 
or mass accretion rates
(e.g. Poutanen et al. 2007; Urquhart and Soria 2016; {Feng et al. 2016}
and reference therein).
{To some extent, this scenario is similar to the unification scenario
proposed for active galactic nuclei (e.g., Elvis 2000, and references therein)}.
The core of our work is to test this model through the 
study of a bright and well isolated transitional object, 
which looks like a ULX when the wind is not optically thick
and then {shows some ULS signatures} when the wind thickens: NGC 55 ULX.

The study of winds in ULXs and ULSs is now possible
after the recent discovery (Pinto et al. 2016, hereafter Paper\,I) of
rest-frame emission and blue-shifted absorption lines in NGC 1313 X-1 
and NGC 5408 X-1 ULXs, 
with the high-resolution RGS gratings aboard XMM-\textit{Newton}. 
{Similar features were also found in NGC 6946 X-1, albeit at lower significance.}
The detections confirm the presence of powerful, relativistic ($\sim0.2c$),
winds in these sources and by extension in several other ULXs 
with similar spectral residuals 
(e.g., Stobbart et al. 2006, Sutton et al. 2015, Middleton et al. 2015b). 
The high-ionization Fe\,K part of the ultrafast outflow in
stacked EPIC and NuSTAR spectra of NGC 1313 X-1 was also detected (Walton et al. 2016a).

The timing properties of ULXs provide an independent diagnostic of the accretion process.  
Heil \& Vaughan (2010) first detected a linear correlation between variability amplitude 
and flux in the ultraluminous X-ray source NGC 5408 X-1, similar to that found in 
Galactic X-ray binaries and active galactic nuclei 
(e.g., Uttley \& McHardy 2001 and Vaughan et al. 2003).  
Heil \& Vaughan (2010) also found evidence of time delays between soft and hard energy bands, 
with the softer bands delayed with respect to the hard at frequencies of $\sim 20$ mHz.  
These time delays were later confirmed by De Marco et al. (2013)
in several newer XMM-\textit{Newton} observations. 
Delays between spectral components may provide some further clues on the geometry 
of the system and the different physical emission mechanisms 
in ULX and ULS accretion disks. 
They could indeed be related to the different location of the soft/hard X-ray emitting 
regions.

This paper is structured as follows. 
In Sect.\,\ref{sec:source} we report some well known 
characteristics of NGC 55 ULX
that motivated us to search for evidence of winds.
We present the data in Sect.\,\ref{sec:data} 
and a detailed spectral modeling in Sect.\,\ref{sec:spectral_modeling}. 
Covariance and lag spectra of the source are shown in Sect.\,\ref{sec:timing}. 
We discuss the results and provide some insights on future missions in Sect.\,\ref{sec:discussion} 
and give our conclusions in Sect.\,\ref{sec:conclusion}.
More technical detail on our analysis is reported in Appendix\,\ref{sec:appendix}.

\section[]{The ULX$-$ULS hybrid in NGC 55}
\label{sec:source}

NGC 55 ULX is the brightest X-ray source in the nearby Magellanic-type 
galaxy NGC 55 (NED average distance $1.94$\,Mpc\footnote{https://ned.ipac.caltech.edu/}) 
with an X-ray luminosity peak of $\sim2\times10^{39}$ erg/s
(see Table\,\ref{table:log} and Fig.\,\ref{Fig:ngc55_epic_map}).
The X-ray light curve exhibits a variety of features including
sharp drops and 100s seconds dips. During the dips most
of the source flux is quenched in the 2.0--4.5 keV band.
Stobbart et al. (2004) proposed that the accretion disk is viewed close 
to edge-on and that, during dips, orbiting clumps of obscuring material 
enter the line of sight and cause significant blocking or 
scattering of the hard thermal X-rays emitted from the inner disk.

The EPIC CCD spectra of NGC 55 ULX 
can be modelled with a broad-band component
from the innermost region (either as a disk blackbody or as Comptonization)
plus the standard T $\sim$ 0.15 keV blackbody, 
which we see in most ULSs and the softest ULXs and
probably originates from reprocessed emission,
such as Compton down-scattered emission,
and intrinsic disk emission
(e.g., Sutton et al. 2013 and Pintore et al. 2015).
The spectrum is very soft (slope $\Gamma>4$, if modelled with a powerlaw)
similar to ULSs, but there is significant flux and a bright hard tail
above 1\,keV (see Sect.\,\ref{sec:epic_continuum}),
which is stronger than typical ULSs (e.g., Urquhart and Soria 2016).
 
\begin{figure}
  \includegraphics[width=0.65\columnwidth,angle=-90,bb=80 55 536 725]{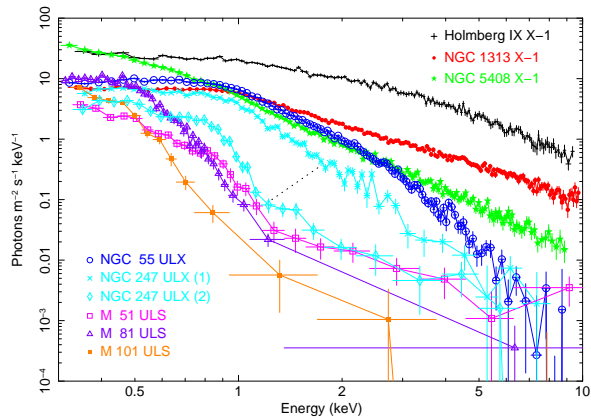}
  \centering
   \caption{ULX-to-ULS spectral states sequence. A dotted line connects the standard ULS states
   with the soft ULX states of NGC 247 and NGC 55 ULXs. For more detail see text.} 
            \label{Fig:ULX_sequence}
\end{figure}

{In Fig.\,\ref{Fig:ULX_sequence} we show the observed EPIC-pn spectrum of
NGC 55 ULX (XMM-\textit{Newton} observation 0655050101) with a sequence of ULX spectra,
from harder to softer: Holmberg IX X-1 (0693851801), NGC 1313 X-1 (0693851201),
NGC 5408 X-1 (0653380501), NGC 247 ULX (0728190101 and 0601010101), M 51 ULS (0303420201),
M 81 ULS (Chandra observation 735), and M 101 ULS (0212480201).}

{NGC\,247 ULX is a very interesting source which oscillate around the boundary between the 
ULS and the soft ULX regime (see also, e.g., Feng et al. 2016). 
Unfortunately, no deep grating observations are available
for this object and for ULSs in general (see Sect.\,\ref{sec:next_decade}).}
{However, the high-flux spectrum of NGC 247 has a similar shape to that 
of NGC 55 ULX with both sources showing a turnover above 1\,keV and,
more importantly, is superimposable with the spectrum of the low-flux state of NGC 55 ULX 
(see, e.g., Fig.\,\ref{Fig:epic_flux_resolved_spectra}).
NGC 55 ULX has never shown a classical ULS spectrum during the XMM-\textit{Newton} observations
because the fraction of the energy carried by the photons
above 1.5 keV has been above 15\%, which is higher than the 10\% limit
commonly used to identify supersoft sources (see, e.g., Di Stefano and Kong 2013),
but its similarity to NGC 247 ULX, which sometimes looks like a ULS, and 
its soft spectral shape, which just fits in between the classical ULXs
and ULSs, are crucial.}

{There is a whole on-axis XMM-\textit{Newton} orbit ($\sim120$\,ks) of NGC 55 ULX }
plus other shorter exposures which provide a unique workbench to study the transition
between the two phases. The search for a wind in this object 
is highly motivated by the detection of soft X-ray residuals 
in the EPIC spectra of NGC 55 ULX that are similar in both shape
and energy ($\sim1$\,keV) to those 
that were observed and resolved in other ULXs 
(see, e.g., Middleton et al. 2015b and Paper\,I).

\begin{table}
\caption{XMM-\textit{Newton} observations used in this paper.}  
\label{table:log}      
\renewcommand{\arraystretch}{1.1}
 \small\addtolength{\tabcolsep}{-1pt}
 
\scalebox{1}{%
\begin{tabular}{c c c c c c c c c c}     
\hline  
ID                     &  Date      & t\,$_{\rm pn}^{(a)}$   & t\,$_{\rm RGS1}^{(a)}$  & t\,$_{\rm RGS2}^{(a)}$ & $L_X^{(b)}$    \\
                       &            & (ks)                   & (ks)                    & (ks)                   &                \\
\hline                                                                                                                    
0028740201             & 2001-11-14 &   27.2                 &   $-$                   &   $-$                  &  2.07          \\
0655050101             & 2010-05-24 &   95.3                 &   119.9                 &   119.5                &  1.30          \\
\hline                
\end{tabular}}

$^{(a)}$ EPIC-pn/RGS net exposure time. 
$^{(b)}$ $0.3-10$\,keV luminosity in $10^{39}$ erg/s assuming a distance of 1.94\,Mpc.
No significant RGS data were available during the first (off-axis) observation.
\end{table}
 
\begin{figure}
  \includegraphics[width=0.95\columnwidth]{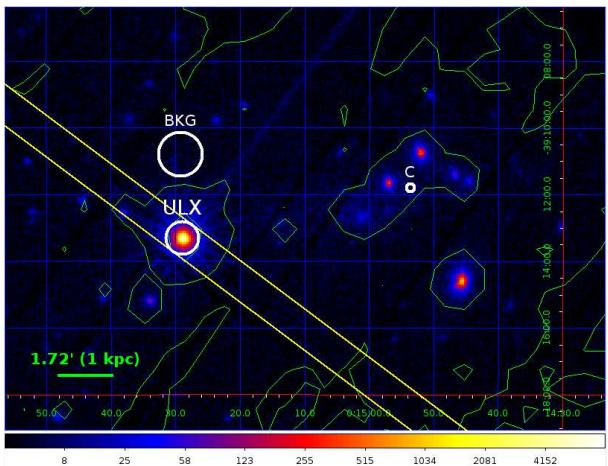}
  \centering
   \caption{NGC 55 ULX EPIC/pn-mos stacked image. 
            The yellow lines show the extraction region for the RGS spectrum.
            The bigger larger white circle is the EPIC background region.
            The smallest circle in the right indicates the NGC 55 galactic center.} 
            \label{Fig:ngc55_epic_map}
\end{figure}

\section[]{The data}
\label{sec:data}

In this work we utilize data from the broadband {($0.3-10$\,keV)}
EPIC-pn CCD spectrometer (Turner et al. 2001)
and the high-resolution Reflection Grating Spectrometer (RGS, {$0.35-1.77$\,keV}) 
aboard XMM-\textit{Newton} (den Herder et al. 2001)
in order to constrain the spectral shape of the source 
and to search for evidence of a wind.

The XMM-\textit{Newton} observations used in this paper 
are reported in Table\,\ref{table:log}.
We decided not to use another available observation (ID=0028740101)
because it had the ULX near the edge of the field of view 
with a PSF highly distorted and affected by the background.
The combination of low flux and short exposure 
with corresponding limited statistics
makes it not useful for our analysis.
The exposure with ID 0655050101 is the only one to provide significant RGS 
science data because the other exposures are off-axis
with the ULX outside the RGS ($\pm2.5'$) field of view.

We reduced the data with the latest XMM-\textit{Newton}/ 
SAS v15.0.0 
(CALDB available on September, 2016) and
corrected for contamination from soft-proton flares.
We extracted EPIC MOS 1-2 and pn images in the 0.5--3\,keV energy range,
which includes the vast majority of NGC 55 ULX photons,
and stacked them with the \textit{emosaic} task
(see Fig.\,\ref{Fig:ngc55_epic_map}).
We also extracted EPIC-pn spectra from within a circular region of 1 
arcmin diameter centred on the emission peak. 
The background spectra were extracted from within a slightly larger
circle in a nearby region on the same chip, 
but away from bright sources and the readout direction
(see Fig.\,\ref{Fig:ngc55_epic_map}).
We also make sure that the background region was outside
the copper emission ring (Lumb et al. 2002).
{We also tested three other background regions for the EPIC-pn
spectrum to confirm that the background does not produce
instrumental features.}
There was very little solar flaring during the observations.

We extracted the first-order RGS spectra in a cross-dispersion region 
of 1 arcmin width, centred on the emission peak 
and the background spectra by selecting photons
beyond the 98\% of the source point-spread-function and check 
for consistency with blank field observations. 
For more detail on the RGS background see Appendix\,\ref{sec:rgs_background}.
We stacked the RGS 1 and 2 (for displaying purposes only,
using the method described in Paper\,I). 
The RGS spectrum of NGC 55 ULX shows very interesting emission-like features
and a sharp drop near 11\,{\AA} very similar to that one observed in 
EPIC spectra of ultraluminous supersoft sources (e.g., Urquhart and Soria 2016).


\section{Spectral analysis}
\label{sec:spectral_modeling}

\subsection{Broadband EPIC spectrum}
\label{sec:epic_continuum}

We fitted the EPIC-pn spectrum with the new SPEX code{\footnote{http://www.sron.nl/spex}}
v3.02 to constrain the broad-band continuum. Each of MOS 1 and 2 cameras 
has 3--4 times less effective area than pn, which alone contains 
the vast majority of the counts.
The pn spectrum was sampled in channels equal to 1/3 of the spectral resolution
for optimal binning and to avoid over-sampling. 
{We grouped the pn spectra in bins with at least 25 counts and use $\chi^2$ statistics.
Throughout the paper we adopt 1\,$\sigma$ error bars.}

 
We focused on the pn spectrum of the second (0655050101) exposure to determine
the best fitting continuum model because it has twice as many counts as the first observation,
shows less variability, and its continuum determination 
is crucial for following high-resolution analysis. 
We tested two alternative models to describe the broadband EPIC-pn spectrum of NGC 55:
a combination of blackbody (\textit{bb}) plus powerlaw (\textit{po}) 
and a combination of blackbody (\textit{bb})
plus disk blackbody modified by coherent Compton scattering 
(\textit{mbb}, see SPEX manual for more detail). 
All emission components were corrected by absorption due to the foreground interstellar
medium and circumstellar medium (\textit{hot} model in SPEX with low temperature 
$0.5$\,eV, see e.g. Pinto et al. 2012).

\begin{table}
\caption{Constraints on the broadband continuum model.}  
\label{table:continuum}      
\renewcommand{\arraystretch}{1.3}
 \small\addtolength{\tabcolsep}{-0pt}
 
\scalebox{1.0}{%
\begin{tabular}{c c c c c c c c c c}     
\hline  
Parameter            &  BB + PO           &  BB + MBB          \\  
\hline                                                                                      
$L_{X\,po}$ (erg/s)  & 1.3$\times10^{40}$  &  --                \\  
$L_{X\,bb}$ (erg/s)  & 1.9$\times10^{38}$  & 7.4$\times10^{38}$ \\  
$L_{X\,mbb}$ (erg/s) &  --                 & 5.5$\times10^{38}$ \\  
$\Gamma $            &  $4.76 \pm 0.10$    & $-$                \\  
$k\,T_{bb}$ (keV)    &  $0.52 \pm 0.09$    & $0.163\pm0.003$    \\  
$k\,T_{mbb}$ (keV)   &  $-$                & $0.69\pm0.01$      \\  
$N_{\rm H}\,(10^{21} {\rm cm}^{-2})$       &  $5.10\pm0.14$  &  $2.07\pm0.08$ \\  
$\chi^2/d.o.f.$      &  332/99             & 191/99             \\  
\hline                                                                                                                
\end{tabular}}

EPIC-pn$_{\,(0655050101)}$ spectrum was used for its high S/N ratio.
$L_{X\,(0.3-10\,\rm keV)}$ luminosities are calculated assuming a 
distance of 1.94\,Mpc and are corrected for absorption (or de-absorbed).
Flux percentages are measured between $0.3-10$\,keV. 
\end{table}

\begin{figure}
  \includegraphics[width=1.325\columnwidth, angle=-90, bb=65 50 565 425]{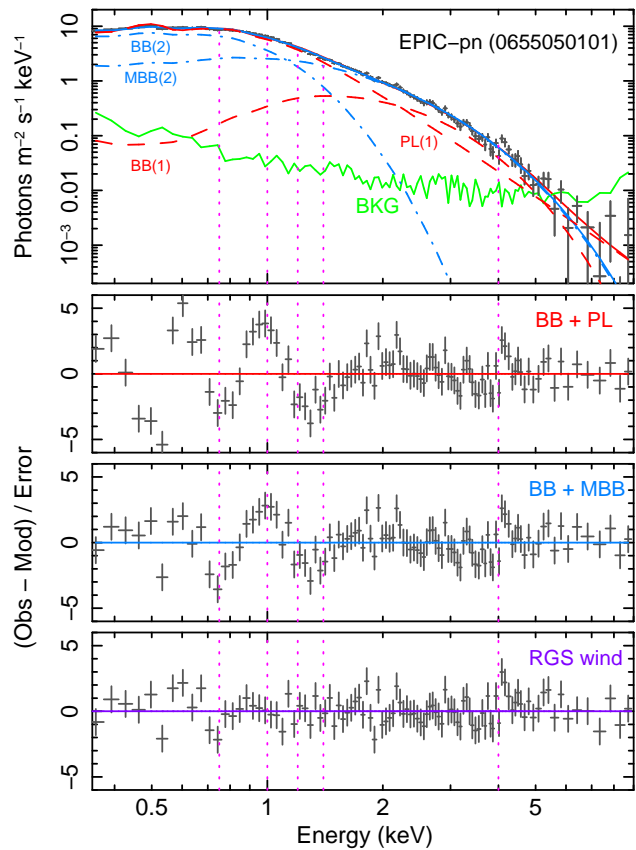}
   \caption{NGC 55 ULX EPIC-pn background-subtracted spectrum with 
         two alternative continuum models (see Table\,\ref{table:continuum}).
         The dotted lines show interesting features that do not depend
         on the adopted continuum model, some of which can be resolved in the soft X-ray 
         high-resolution RGS spectrum 
         (see Fig.\,\ref{Fig:rgs_spectrum}). 
         The dotted lines show the model components for Model 1 (BB+PL)
         and Model 2 (BB+MBB); for more detail see Table\,\ref{table:continuum}.
         The bottom panel show the EPIC-pn residuals adopting the wind model
         from the RGS high-resolution spectrum (see Sect.\,\ref{sec:rgs_model_onto_epic_data}).} 
            \label{Fig:epic_continuum}
\end{figure}

The double blackbody model (\textit{bb} + \textit{mbb}) provides a better fit than the power-law model  
(see Table\,\ref{table:continuum}) in agreement with the
two-component model consisting of a blackbody (for the
soft component) and a multicolour accretion disk (for the hard component)
of Pintore et al. (2015) for \textit{Chandra} and \textit{Swift}
observations of NGC 55 ULX. 
Moreover, the power-law model would imply a very high luminosity
since the deabsorbed model diverges at low energies.

\begin{figure}
  \includegraphics[width=0.95\columnwidth]{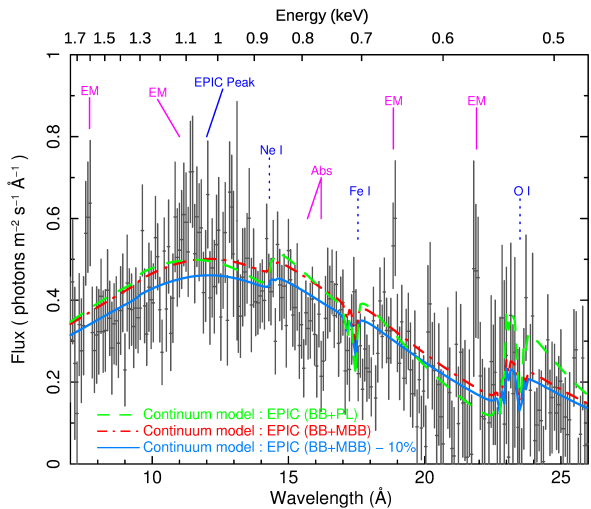}
   \caption{RGS stacked spectrum with overplotted the EPIC best-fit continuum
         model (see Table\,\ref{table:continuum}). The solid blue line shows the RGS fitted continuum
         with free normalization as only free parameter.
         The wavelengths of expected absorption lines produced by the Galactic absorber 
         (Pinto et al. 2013)
         and the $\sim1$\,keV peak seen in the EPIC spectrum are shown by blue dashed
         lines. Some clear emission- (EM) and absorption-like (Abs) spectral features 
         are marked with solid magenta lines.} 
            \label{Fig:rgs_spectrum}
\end{figure}

{Of course, we cannot exclude one or the other continuum model 
according to the value of the $\chi^2$ only. A good exercise consists 
in determining the column density, $N_{\rm H}$ in a different way or using other 
observations and compare it to that measured with the blackbody and the power-law
models. Pintore et al. (2015) have shown that the $N_{\rm H}$ of NGC 55 ULX 
varies over a large range, $1-6 \times 10^{21} {\rm cm}^{-2}$, which is always 
an order of magnitude above the Galactic value ($1-2 \times 10^{20} {\rm cm}^{-2}$).
Interestingly, if we fit the EPIC data with the power-law model,
but excluding the hard tail above 2 keV, we obtain $4.1\times10^{21} {\rm cm}^{-2}$.
If instead we only fit the EPIC spectrum between 0.4 and 1 keV we 
obtain $2.7\times10^{21} {\rm cm}^{-2}$, which may suggest
that the broadband power-law fit overestimates the column density.
We have also tested a Comptonization continuum (\textit{comt} model in SPEX 
with $T_{\rm seed}=0.13\pm0.01, T_{\rm e}=0.83\pm0.03, \, {\rm and} \; \tau=6.8\pm0.3$)
over the 0.3--10\,keV and obtained $2.1 \pm 0.1\times10^{21} {\rm cm}^{-2}$.
Following the procedure of Pinto et al. (2013), 
adopting Solar abundances and a power-law continuum, 
we have also measured the $N_{\rm H}$ through the depth of the absorption edges 
of the most abundant interstellar neutral species in the RGS spectrum:
neon (13.4\,{\AA}), iron (17.4\,{\AA}), and oxygen (23.0\,{\AA}),
see Fig.\,\ref{Fig:rgs_spectrum}.
We obtained $N_{\rm H}=1.5\pm0.5 \times 10^{21} {\rm cm}^{-2}$,
without significant differences whether we fit the whole RGS spectrum or edge-by-edge.
This also prefers the solution obtained with the blackbody model.}

The harder multicolour component is seen in most ULXs,
although its temperature is typically 1.5--2.5 keV 
(hence the origin of the characteristic downturn at $\sim$ 5--6 keV).
In the softest states of ULXs {(e.g. in NGC 55 ULX)} 
and in {classical} ULSs can be as low as 0.7 keV. 
It is thought to consist of emission produced by the {inner} accretion disk 
highly distorted by interaction with hot electrons in the inner regions 
(up-scattering) and by down-scattering with cool electrons in the wind
(e.g. Gladstone et al. 2009, Middleton et al. 2011a, Middleton et al. 2015a).
The soft ($\sim0.2$\,keV) component is likely to be intrinsic to 
regions above the {outer} disk, where a radiatively-driven wind is expected 
to be launched at accretion rates comparable or higher than the Eddington limit 
(Poutanen et al. 2007). 
Our broadband spectral fits are consistent 
with NGC 55 ULX being a soft ultraluminous (SUL) X-ray 
source according to the classification of Sutton et al. (2013).

The first, shorter, observation also favours a multi-blackbody emission model.
The EPIC spectrum in 2001 basically exhibits hotter blackbody components
(see below and Table\,\ref{table:epic_flux_resolved_spectra}). 
Interestingly, we found that the emitting area of the soft ($\sim0.2$\,keV) blackbody 
component increases by a factor 2 from the brighter (2001) 
to the fainter (2010) observation in agreement with the study of ULSs
of Urquhart and Soria (2016) and Feng et al. (2016). This is expected
if the expansion of the photosphere and the decrease of the temperature
shift the peak of the spectrum from X-ray towards the far-UV.

As previously mentioned, ULSs show a faint hard X-ray tail ($>$ 1.5 keV) 
whose origin is not quite understood. 
It is thought to be produced in the inner regions 
similar to ULXs and characterized by either Bremsstrahlung 
or Comptonization emission (see, e.g., Urquhart and Soria 2016). 
As a test, we have re-fitted the NGC 55 ULX EPIC spectrum 
by substituting the \textit{mbb} component with a rest-frame Bremsstrahlung 
or collisionally-ionized emission model ($cie$ model in SPEX).
The \textit{bb + brems} fit is bad ($\chi^2_{\nu} = 13.8$).
We also tried to get an upper limit on the emission of a putative Bremsstrahlung
component adding it on top of the best-fit \textit{bb + mbb} model.
The maximum flux a Bremsstrahlung component can have is 10 times smaller 
than the flux of the \textit{mbb} component. 
The Bremsstrahlung temperature was however highly unconstrained.
Therefore, if the hard X-ray tail seen in ULSs is similar 
to the one that is well constrained in NGC 55 ULX, 
then a disk origin would be favoured.

Although the double blackbody model provides a better fit,
it is still statistically rejected.
The presence of strong residuals around 1\,keV 
shows an important similarity with the residuals seen in several 
ULXs with long exposures (emission peak at 1\,keV and absorption-like
features on both sides).
Middleton et al. (2014, 2015b) interpreted them as either due
to emission by collisionally-ionized gas or to absorption
by outflowing photoionized gas and were identified in Paper\,I.
This motivated our investigation with the RGS.

{The knowledge of the exact structure of the spectral continuum
(either a double black body or power-law or Comptonization)
does not strongly affect the following analysis, which focuses on the 
search for narrow spectral features and broadband spectral
variations.}

\subsection{The high-resolution RGS spectrum}
\label{sec:rgs_spectra}

\subsubsection{RGS spectral continuum}
\label{sec:rgs_continuum}

Our analysis focuses on the $7-27$\,{\AA} first order 
RGS 1 and 2 spectra from the long, 
on-axis, observation 2 (0655050101).
Below 7\,{\AA} the effective area is not well calibrated
and above 27\,{\AA} the count rate is too low.
For the RGS analysis we have to use C-statistics 
because of the Poisson statistics
and the need for high spectral resolution.
The spectrum was binned in channels equal to 1/3 of  
the spectral resolution for the optimal binning 
(see Kaastra and Bleeker 2016).

We plot the EPIC continuum models on top of the RGS data in Fig.\,\ref{Fig:rgs_spectrum}.
As for EPIC, the double blackbody spectral model describes the RGS spectrum better 
than the blackbody+powerlaw combination, although RGS does not have the same sensitivity.
Therefore, the choice of the continuum will not have a {significant} effect on our results.
{We fit the EPIC model to the RGS spectrum leaving 
only the overall normalization as a free parameter
and obtained a comparable fit 
with the overall flux lower by about 10 percent compared to the EPIC continuum model 
(see the solid blue line in Fig.\,\ref{Fig:rgs_spectrum}).
This RGS continuum model is used in the following analysis}.

One of the major issues in modelling the RGS spectra of faint sources, such as ULXs,
is the background. However, NGC 55 ULX is significantly brighter than the background 
between 7--25\,{\AA}. Moreover the source is reasonably isolated;
the RGS slit does not encounter any other bright object (see Fig.\,\ref{Fig:ngc55_epic_map})
and the background spectrum is rather featureless.
For more detail see Appendix\,\ref{sec:rgs_background}.

\subsubsection{Search for spectral features}
\label{sec:line_detection}

\begin{figure}
  \includegraphics[width=0.99\columnwidth]{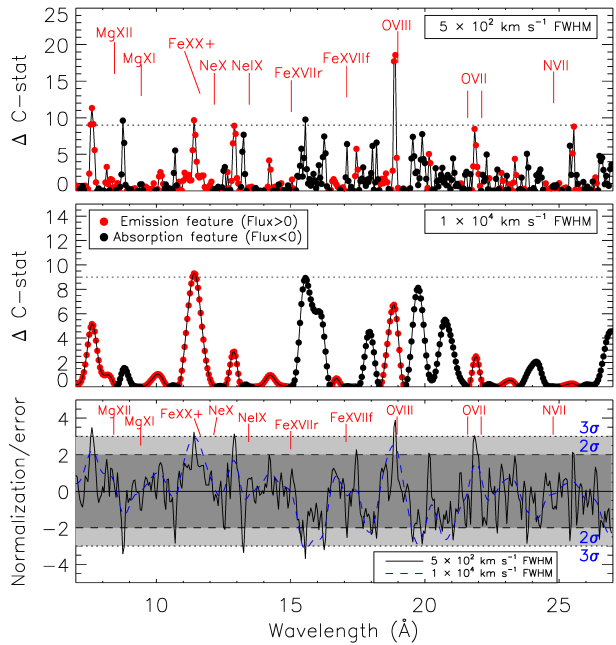}
   \caption{$\Delta$\,C-statistics 
         obtained by fitting a Gaussian spanning the $7-27$\,{\AA} wavelength
         range with increments of 0.05\,{\AA}.
         $\Delta$\,C = 2.71, 4.00, and 9.00 refer
         to 1.64$\sigma$ (90\%), 2$\sigma$ (95.45\%), 
         and 3$\sigma$ (99.73\%) confidence levels.
         We tested the effects of using two different linewidths: 500 km s$^{-1}$ 
         ($\sim$ RGS resolution, top panel) and 10,000 km s$^{-1}$ (medium panel). 
         Red and black points refer to emission and absorption lines, respectively. 
         The ratio between the normalization of the gaussian and its error
         is shown in the bottom panel.
         The rest-frame wavelengths of some relevant transitions are labelled.
         More detail on the line significance is report in the text.} 
            \label{Fig:line_detection}
\end{figure}

Following the approach used in Paper\,I,
we searched for spectral features on top of the spectral continuum
by fitting a Gaussian spanning the $7-27$\,{\AA} wavelength range with 
increments of 0.05\,{\AA} and calculated the $\Delta$\,C-statistics.
We tested a few different linewidths: 500, 1000, 5000, and 10,000 km s$^{-1}$. 
We show the results obtained with 500 km s$^{-1}$ ($\sim$ RGS resolution) 
and 10,000 km s$^{-1}$ (a case of relativistic broadening) 
in Fig.\,\ref{Fig:line_detection}.
The black points refer to lines with negative normalizations, i.e. absorption lines.
We found several narrow emission features
and some evidence for absorption. 
The emission-like features are near some
of the strongest transitions commonly observed in X-ray plasmas 
and are the same that we have detected in our previous
work on NGC 1313 ULX-1, NGC 5408 ULX-1, and NGC 6946 ULX-1.
The RGS spectra amplify and resolve the residuals 
previously detected in the EPIC spectra 
(Middleton et al. 2015b). 

If we identify the emission features with 
some of the strongest transitions in this energy band 
-- {\mgxii} (8.42\,{\AA}), {\fexxii-iii} (11.75\,{\AA}), 
{\nex} (12.135\,{\AA}), {\neix} (13.45\,{\AA}), 
{\oviii} (18.97\,{\AA}) and {\ovii} (21.6\,{\AA}, 22.1\,{\AA}) --
then they would require blueshift, which seems to be larger 
for higher ionization states.
For instance, the spike around 7.5\,{\AA} is likely produced 
by blueshifted {\mgxii}; the one seen around 11.4\,{\AA}
could be a blend of blueshifted {\nex} and high Fe ions.
The 21.8\,{\AA} emission-like feature could be a 
blueshifted {\ovii} 22.1\,{\AA} forbidden line.
An {\ovii} forbidden line (22.1\,{\AA}) stronger 
than the resonant line (21.6\,{\AA})
is a clear indication of photoionization (e.g. Porquet and Dubau 2000).
The high ionization ions, e.g. {\mgxii}, {\fexxii-iii} and {\neix}
also show strong resonant lines suggesting possible collisional ionization 
and, therefore, a complex wind ionization/temperature structure.
{In the identification process we assumed that the lines 
were produced by the strongest transitions in the X-ray band and 
within $\pm0.2c$ from their rest-frame wavelengths.} 
{We notice that the blueshifts of the strongest {\ovii} and {\oviii} emission lines 
are consistent within 1\,$\sigma$. The same applies to the {\mgxii} and {\nex} emission lines.}

{In Appendix\,\ref{sec:rgs_spectra_blazars} we show that the features detected in
the RGS spectrum of NGC 55 ULX are intrinsic to the sources and are not of 
intrumental origin by applying the same line-search technique to the spectra of 
five active galactic nuclei with comparable statistics.}

\subsubsection{Monte Carlo simulations}
\label{sec:monte_carlo_simulations}

The $\Delta$\,C-statistics of each feature provides a crude estimate
of the significance but does not take into account the look-elsewhere-effect,
which is due to the size of the parameter space used such as the number of velocity bins used,
and other eventual effects due to the data quality. 
We therefore performed Monte Carlo simulations to estimate the significance
of the strongest features in the RGS spectrum such as those 
found at 7.5\,{\AA}, 11.4\,{\AA}, 15.55\,{\AA}, 18.9\,{\AA}, and 21.85\,{\AA}.
Adopting our best-fit spectral continuum model as template 
(Table\,\ref{table:continuum}, second column),
we simulated 10,000 RGS 1 and 2 spectra {(accounting for the uncertainties 
on the continuum parameters). We then added the line as best-fitted in our line grid 
to the continuum model and calculated the changes in the C-statistics}
(see Fig.\,\ref{Fig:line_detection}). 
For each line, we then computed the number of occurrences where its addition
to the continuum-simulated spectra provided an improvement 
(or $\Delta$\,C-statistics, {{with the position of the line 
free to vary within $\pm0.2c$} and the continuum parameters also free to vary})
larger than that one provided by the line as fitted to the data.
{This is basically equivalent to create a mock spectrum from the 
continuum model and search for lines in this spectrum using the same technique
adopted in Sect.\,\ref{sec:line_detection}}.
The $p$-values and the corresponding confidence intervals (C.I.) and sigmas
are reported in Table\,\ref{table:monte_carlo_simulations}.
This provides much more conservative limits on the detection.
As expected, the blueshifted {\oviii} line ($\lambda_0=$18.97\,{\AA}, 
$\lambda_{\rm obs}\sim$18.90\,{\AA}) 
is detected well above 3\,$\sigma$, but all the strongest features 
are still detected above the $99\%$ C.I. 
each, even accounting for the look-elsewhere effect.

\begin{figure*}
  \includegraphics[width=1.85\columnwidth]{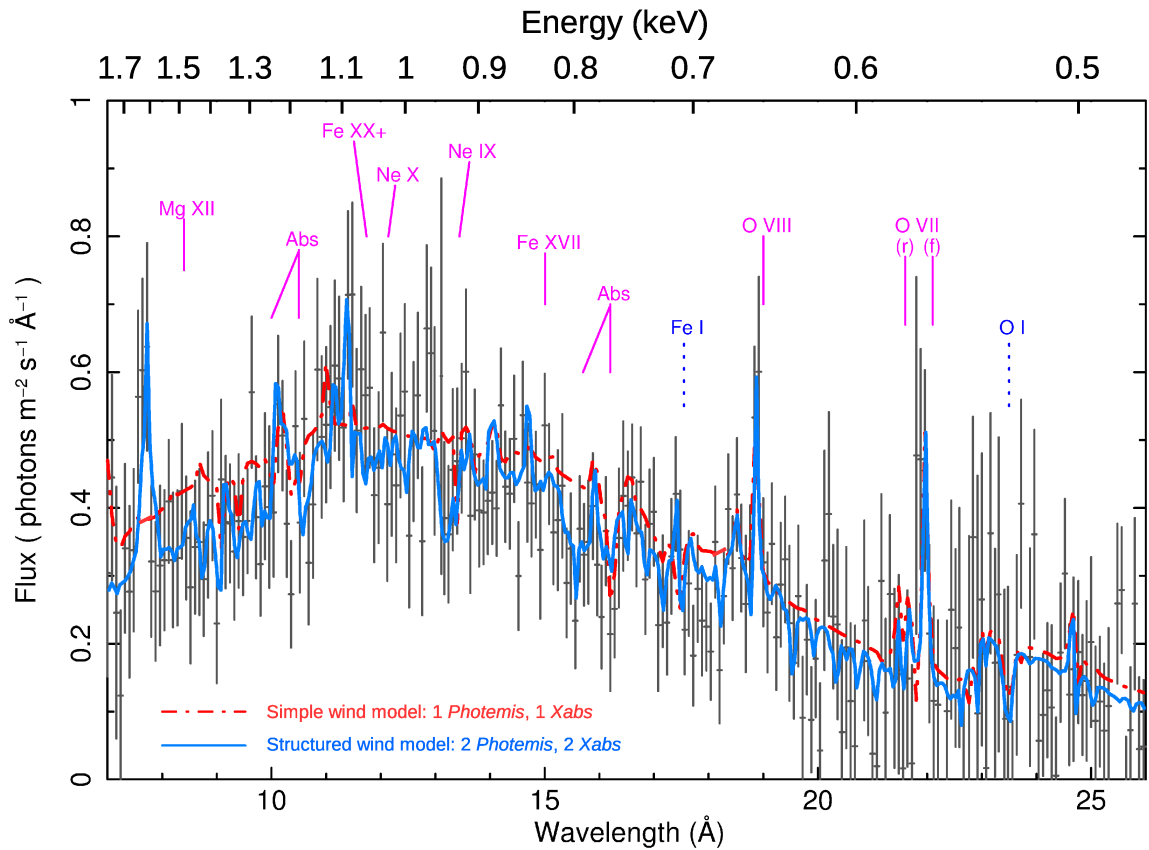}
   \caption{RGS spectrum with two alternative 
         wind models on top of the best-fit EPIC continuum:
         the red line has one photoionized outflowing emission ($0.01c$) and one
         absorption ($0.16c$) components;
         the blue line has two photoionized emission components ($0.01c$ and $0.08c$)
         and two absorbers ($0.06c$ and $0.20c$, see
         Table\,\ref{table:improvements}).
         Rest-frame wavelengths of relevant transitions are labelled.
         Note the line shift.} 
            \label{Fig:rgs_spectrum_fit}
\vspace{-0.3cm}
\end{figure*}

{Although this procedure does not follow exactly all the steps 
adopted during the line search performed in Sect.\,\ref{sec:line_detection},
we acknowledge that it is the best that can be done with the current data.
The combined detection of two pairs of strong lines sharing the velocity
is certainly encouraging, see also {\mgxii}--{\nex} and {\ovii}--{\oviii} lines 
in Fig.\,\ref{Fig:rgs_spectrum_fit}, but more data are required to place strong constraints.}

\begin{table}
\caption{NGC 55 ULX XMM/RGS fit : Monte Carlo simulations.}  
\label{table:monte_carlo_simulations}             
\renewcommand{\arraystretch}{1.0}      
 \small\addtolength{\tabcolsep}{-3pt}
 
\scalebox{1}{%
\begin{tabular}{c|c c c c c c c c c}     
\hline
$\lambda$                          & $7.5$\,{\AA} & 8.75\,{\AA} & 11.4\,{\AA} & 15.55\,{\AA} & 18.9\,{\AA} & 21.85\,{\AA} \\
\hline                                                  
Type                               &  E           &  A          &  E          &  A           &  E          &  E           \\
\hline                                                                                      
$p$-value                          & 0.005        & 0.007       & 0.009       & 0.009        & 0.001       & 0.008        \\
 C.I.                              & $99.5\%$     & $99.3\%$    & $99.1\%$    & $99.1\%$     & $99.9\%$    & $99.2\%$     \\
$\sigma$                           & $2.8$        & $2.7$       & $2.6$       & $2.6$        & $3.3$       & $2.65$       \\
\hline
\end{tabular}}

{ Confidence level of the lines detected in the RGS spectrum.
 The $p$-values were computed with 10,000 Monte Carlo simulations
 accounting for the number of trials
 (for more detail see Sect.\,\ref{sec:monte_carlo_simulations}).
 ``E'' and ``A'' indicate emission and absorption lines, respectively.}

 \end{table}

\begin{table*}
\caption{NGC 55 ULX XMM/RGS fit : main wind components.}  
\label{table:improvements}             
\renewcommand{\arraystretch}{1.1}      
 \small\addtolength{\tabcolsep}{1pt}
 
\scalebox{1}{%
\begin{tabular}{c|c c|c c c c c c c}     
\hline
Parameter                          & Slow emission   & Fast emission   & Slow absorption & {Fast absorption} \\
\hline                                                                 
$\log \xi$                         & $1.20 \pm 0.25$ & $2.80 \pm 0.30$ & $0.5 \pm 0.3$   & $3.35 \pm 0.20$ \\
$v_{\rm outflow}$                  & $0.011\pm0.001$ & $0.082\pm0.001$ & $0.058\pm0.004$ & $0.199\pm0.003$ \\
$\Delta\,\chi^2$/$\Delta$\,$C/dof$ & 28/27/3         & 30/30/3         & 14/14/3         & 19/18/3         \\
\hline
\end{tabular}}

{ RGS fits wind components:
  photoionized emission and photoionized absorption (see also Fig.\,\ref{Fig:rgs_spectrum_fit}, 
  {solid blue line}, and Fig.\,\ref{Fig:wind_structure_IDL}).   \\
  The photoionization parameters are in log ($\xi$, erg cm s$^{-1}$). 
  Velocities are in units of light speed $c$.}

 \end{table*}
         
\begin{figure}
  \includegraphics[width=0.975\columnwidth]{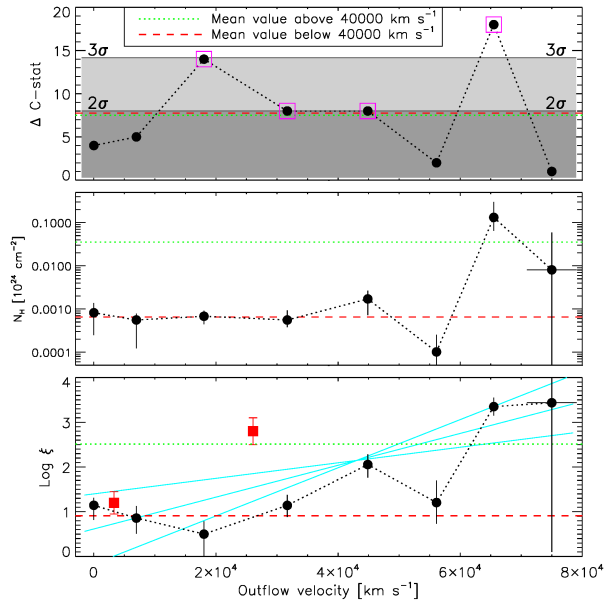}
   \caption{Results of the spectral fits using a grid of 8 photoionized absorbers 
            (see Sect.\,\ref{sec:complex_wind_model}). 
            The $\Delta\,C-$statistics indicate the significance of each absorber 
            (each with three degrees of freedom). 
            The column densities, the ionization parameters,
            and the absolute values of the outflow velocities are also shown.
            The cyan solid lines in the bottom panel show a linear fit with
            the corresponding 3\,$\sigma$ limits.
            The red filled squares indicate the values of the photoionized
            emission components (see Table\,\ref{table:improvements}).
            The transmission of the four absorbers indicated 
            by the magenta open squares
            is shown in Fig.\,\ref{Fig:wind_structure_QDP}.
            The two absorbers with highest $\Delta C$-stat are 
            also used in the multiphase wind model shown in
             Fig.\,\ref{Fig:rgs_spectrum_fit} (blue line).}
            \label{Fig:wind_structure_IDL}
\end{figure}
         
\begin{figure}
  \includegraphics[width=0.95\columnwidth]{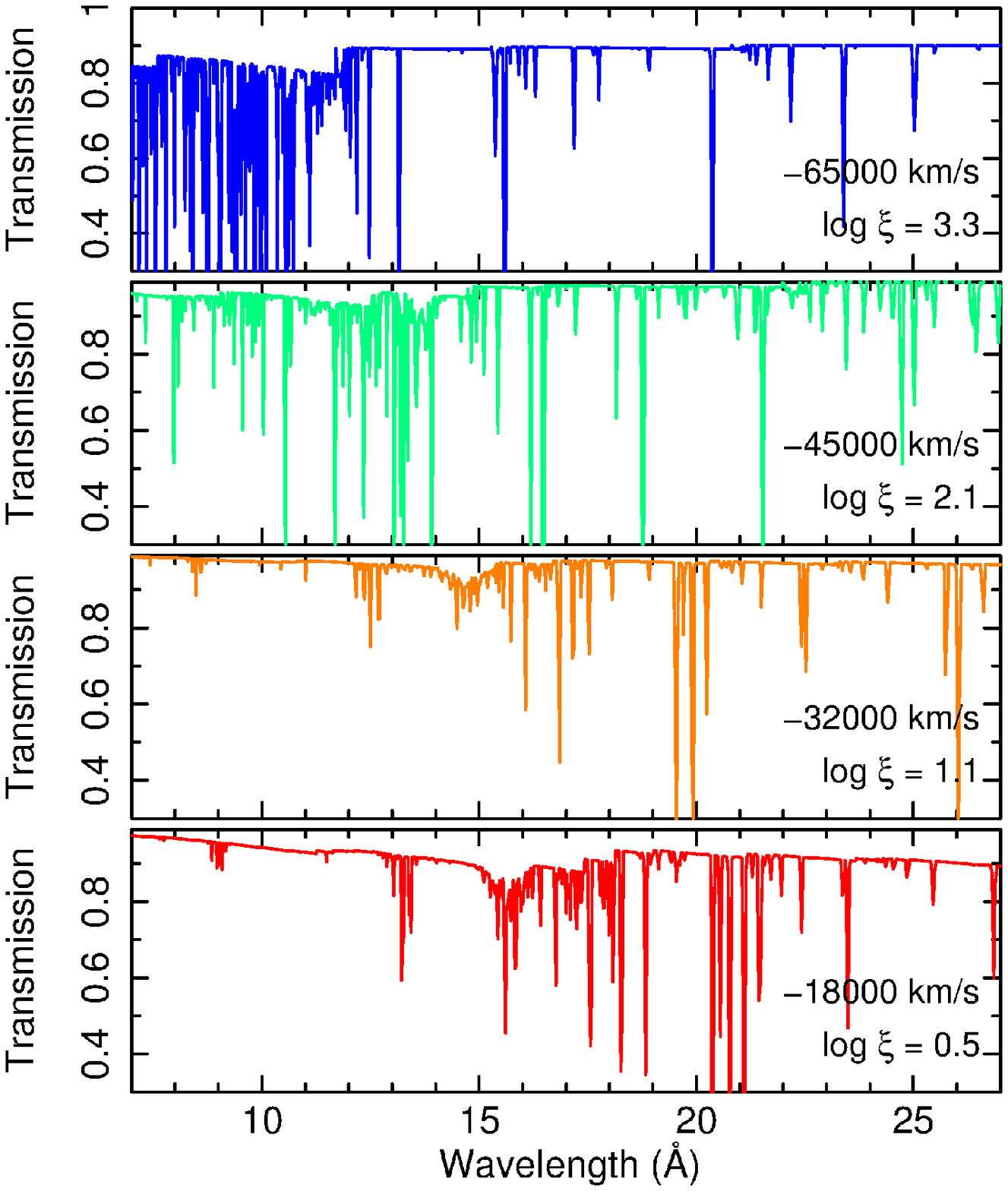}
   \caption{Transmission of the four most relevant absorbers 
            according to a grid of 8 photoionized absorbers 
            (see the magenta open squares Fig.\,\ref{Fig:wind_structure_IDL}).
            The outflow velocities and the ionization states are labelled.
            The fastest absorber significantly reduces the flux below 12\,{\AA}.} 
            \label{Fig:wind_structure_QDP}
\end{figure}
         
\subsubsection{Simple wind model}
\label{sec:simple_wind_model}

We modelled the low-ionization emission lines with a 
photoionized emission model ($photemis$ code imported from 
XSPEC, {which uses XSTAR tables{\footnote{https://heasarc.gsfc.nasa.gov/xstar/docs/html/node106.html}}}).
This model provides a good description of the {\oviii} resonant line, 
the {\ovii} k\,$\alpha$ triplet and 
the {\ovii} k\,$\beta$ line at 18.627\,{\AA} $(\log \xi = 1.2 \pm 0.1)$,
but requires a blueshift of $(0.011\pm0.001)c$, i.e. $\sim$ 3000 km s$^{-1}$.
This model provide a significant improvement to the fit with
$\Delta\,\chi^2$/$\Delta$\,$C/dof$ = 28/27/3
(see also Table\,\ref{table:improvements}).
We notice that SPEX calculates the (blue/red) shift of any
spectral model according to $E' = E / ( 1 - v / c)$,
which does not account for the relativistic corrections.
This was not taken into account in Paper\,I.
Hence, we show all the velocities as defined in SPEX
to allow the readers to reproduce our work,
but we then use the relativistic formula $z = \sqrt{ (1 + v/c) / (1 - v/c) } - 1$
each time we report the blueshift, $z$, of a certain line or model.

As previously done in Paper\,I, we modelled the absorption features 
with the $xabs$ model in SPEX.
This model calculates the transmission of a slab of material, 
where all ionic column densities are
linked through a photoionization model. 
The relevant parameter is the ionization parameter $\xi= L/nr^2$,
with $L$ the source luminosity, $n$ the hydrogen density and $r$ the distance from the ionizing source
(see, e.g., Steenbrugge et al. 2003). 
The equivalent of this model in XSPEC is
$warmabs$\footnote{https://heasarc.gsfc.nasa.gov/xstar/docs/html/node102.html}.
Some absorption-like features can be well described with one photoionized $xabs$ absorber
with a higher ionization parameter $(\log \xi = 3.5 \pm 0.3)$ 
and a relativistic outflow velocity of 
$(0.16\pm0.01)c$, very similar to the $\sim0.18c$ outflow
in NGC 1313 X-1 and NGC 5408 X-1 ($0.2c$ for the non-relativistic correction,
see Paper\,I), see red line in Fig.\,\ref{Fig:rgs_spectrum_fit}.
The absorber reproduces part of the drop below 11\,{\AA} and 
the strong absorption at 16\,{\AA}.
Several features are however missed by this simple wind model
such as the blueshifted {\mgxii} (8.42\,{\AA}) emission line,
part of the {\fexxii-iii} (11.75\,{\AA}) -- {\nex} (12.135\,{\AA}) blend
and some other absorption like features such as the dip at 15\,{\AA}.
This suggests that the wind may be more structured.
Line widths of 100 km s$^{-1}$ were adopted for all 
absorption and emission components.
A different line width does not improve the fit.

\subsubsection{Structured wind model}
\label{sec:complex_wind_model}

\textbf{Multiple emission components --} 
A significant improvement ($\Delta\,\chi^2$/$\Delta$\,$C/dof$ = 30/30/3) 
is obtained if we include another, faster, photoionized emission
with $0.082\pm0.001c$, which provides an excellent
description of the emission features at 7.5\,{\AA} and 11--13\,{\AA} 
(see blue line in Fig.\,\ref{Fig:rgs_spectrum_fit} and Table\,\ref{table:improvements}).
A comparable, slightly worse, fit ($\Delta\,\chi^2$/$\Delta$\,$C/dof$ = 25/25/3)  
is obtained if we instead use a fast collisionally-ionized emission component 
($cie$ in SPEX, $apec$ in XSPEC).
 
\noindent \textbf{Multiple absorption components --} 
The absorption features are less strong and less  significant than the emission lines,
but we have seen that a single component cannot reproduce all the strongest features.
{For instance, the troughs between 8.8--10\,{\AA} and 15--16\,{\AA} (or 1.2--1.4\,keV
and 0.77--0.83, both also seen in EPIC-pn, e.g. Fig.\,\ref{Fig:epic_continuum}).}
However, adding blindly more absorption components may miss some important features.

First, we considered a wind with a continuous distribution of ionization parameters
from 0.2 to 3.8 all at the same velocity. This was done using the $warm$ model in SPEX
which adopts a continuous distribution of 19 absorbers with a $\Delta\xi$ step of 0.2.
This only provided a very small improvement with 
$\Delta\chi^2$ and $\Delta C=26$ for 19 new degrees of freedom 
with respect to the model of continuum plus emission lines. 
We notice however that the best-fit outflow velocity of such multi-ionization component 
is $66000$ km s$^{-1}$, i.e. $0.2c$
similar to NGC 1313 and NGC 5408 ULX-1. 
This motivates a different fitting tactic.

The presence of emission lines at different velocities suggests a structured wind
and, therefore, the absorption lines may also exhibit some complex dynamics.
To search for a trend of the velocity with the ionization state as well as for 
some missing components, we created a grid of photoionized absorbers
with outflow velocities spanning the $0-0.25c$ range 
(8 $xabs$ components in SPEX with velocity steps of about 10,000 km s$^{-1}$ each).
We fit again the RGS data with the EPIC spectral continuum and calculated 
the improvement in the C-statistics with the addition of each component.
There are three degrees of freedom for each absorber: the column density $N_{\rm H}$,
the ionization parameter $\xi$, and the line-of-sight velocity $v$.
The results are shown in Fig.\,\ref{Fig:wind_structure_IDL}.

We found at least two $xabs$ components with high significance: 
$\sim3\sigma$ ($\Delta C$ / $\Delta \chi^2$ / d.o.f. = 14/14/3) 
and $\sim3.5\sigma$ ($\Delta C$ / $\Delta \chi^2$ / d.o.f. = 18/19/3) respectively, 
at low ($\sim0.5$) and high ($\sim3.3$)
ionization parameters. The low-$\xi$ plasma has a low velocity $\sim0.06c$, 
while the high-$\xi$ has is relativistically outflowing $v\sim0.20c$.
There are several tentative ($2\sigma$) detections of material at intermediate velocities,
which were broadly reproduced by the single absorber in the simple model
used in Sect.\,\ref{sec:simple_wind_model}.
Interestingly, the ionization parameter of the absorbers 
seems to show a strong trend with the velocity 
(see Fig.\,\ref{Fig:wind_structure_IDL}, bottom panel).
The effective area of the RGS detector drops dramatically at energies below 1.77\,keV
(or wavelengths above 7\,{\AA}) with a corresponding loss of sensitivity to ionization parameters $\log \xi \gtrsim 4.0$.
This explains the large error bars at the highest velocities.

In Fig.\,\ref{Fig:wind_structure_QDP} we show the transmission of the four 
absorbers with the highest $\Delta C$ (8, 8, 14, and 18), 
which total $\Delta C = 48$ for 12 degrees of freedom). 
This plot is a simple diagnostic tool that shows how the lines shift 
with the higher outflowing velocities and ionization parameters.
We check the effect of a multiphase wind model on the overall spectral fitting by 
fitting again the RGS spectrum with the two most significant photoionized absorbers 
($v\sim0.06c$ and $\sim0.20c$) and the two photoionized emitters discussed above 
on top of the EPIC spectral continuum model.
This structured wind model is shown in Fig.\,\ref{Fig:rgs_spectrum_fit} (blue line)
and {provides $\Delta C = 32$ for 6 degrees of freedom}.
The model fits the RGS spectrum reproducing most narrow emission
and absorption lines such as the drops below 11.5\,{\AA} and between 15--16\,{\AA}.
In Table\,\ref{table:improvements} we report the improvements in the fit 
for these four most significant emission and absorption components.
In Appendix\,\ref{sec:systematics_ionbal} we briefly discuss some systematic effects 
due to the adopted ionization balance.

\subsubsection{Do RGS wind components improve EPIC data fits?}
\label{sec:rgs_model_onto_epic_data}

{A simultaneous fit of RGS and EPIC spectra is not straightforward
due to their different characteristics. EPIC-pn has high count rate but low spectral resolution,
whilst RGS has low count rate but high spectral resolution. Moreover, there are some
cross-calibration uncertainties between them and the energy band of EPIC (0.3--10 keV)
is much wider than that of RGS ($\sim$ 0.35--1.77 keV). 
However, it is a good excercise to test 
the wind components from the RGS spectral modeling onto the EPIC data and confirm 
their existence. Therefore, starting from the EPIC-pn continuum spectral model
in Sect.\,\ref{sec:epic_continuum} (ID 0655050101), we have added each
of the RGS wind components shown in Table\,\ref{table:improvements} on top
of the EPIC double blackbody continuum with only the continuum parameters
free to vary. We remind that we used $\chi^2$ in the EPIC fits. 
The addition of the slow low-$\xi$ emitter has a small effect on the fits 
($\Delta \chi^2 = 10$) because is degenerate with the column density
at the low spectral resolution of EPIC. The fast high-$\xi$ emitter
helds $\Delta \chi^2 = 12$. Interestingly, if we use the collisionally-ionized
RGS solution (see Sect.\,\ref{sec:complex_wind_model}) they provide
a $\Delta \chi^2 = 15$ each. The introduction of the relativistic ($0.2c$) 
photoionized absorber also improves the fits with $\Delta \chi^2 = 27$.
The introduction of these components smears out most of the residuals
that were detected with a continuum-only model (Fig.\,\ref{Fig:epic_continuum}).
The EPIC-pn spectrum provides further support to the results obtained 
with the RGS only. 
}

\subsection{Flux-resolved spectroscopy}
\label{sec:epic_flux_resolved_spectra}

\begin{table*}
\caption{XMM-\textit{Newton}/EPIC broadband continuum models of flux-resolved spectra.}  
\label{table:epic_flux_resolved_spectra}      
\renewcommand{\arraystretch}{1.3}
 \small\addtolength{\tabcolsep}{-2pt}         
 
\scalebox{1}{%
\begin{tabular}{c|c c|c c c c c c c}       
\hline  
Component/parameter                        & OBS 1 high      & OBS 1 low       &  OBS 2 high     &  OBS 2 medium   &  OBS 2 low      \\  
Flux\,$(10^{-12} {\rm erg\, s}^{-1} {\rm cm}^{-2})$  & 6.74  & 5.03            &  3.56           &  3.06           &  2.57           \\  
$L_X\,(10^{39} {\rm erg\, s}^{-1})$        &  $3.04$         &  $2.27$         &  $1.60$         &  $1.38$         &  $1.16$         \\  
\hline                                                                                                                                                             
$N_{\rm H}\,(10^{21} {\rm cm}^{-2})$       & \multicolumn{2}{c}{$1.45\pm0.09$} &         \multicolumn{3}{c}{$2.07\pm0.08$}           \\  
$L_X\,(10^{39}\,{\rm erg\, s}^{-1})_{bb}$  &  $1.23\pm0.32$  &  $1.21\pm0.28$  &  $0.91\pm0.19$  &  $0.80\pm0.16$  &  $0.67\pm0.14$  \\  
$R\,{(\rm km})_{bb}$                       &  $2320\pm620$   &  $2310\pm530$   &  $3160\pm650$   &  $2960\pm610$   &  $2710\pm560$   \\  
$k T\,{(\rm keV})_{bb}$                    &\multicolumn{2}{c}{$0.205\pm0.006$}&         \multicolumn{3}{c}{$0.163\pm0.003$}         \\  
$L_X\,(10^{39}\,{\rm erg\, s}^{-1})_{mbb}$ &  $1.80\pm0.24$  &  $1.05\pm0.19$  &  $0.69\pm0.08$  &  $0.58\pm0.07$  &  $0.49\pm0.06$  \\  
$R\,{(\rm km})_{mbb}$                      &  $136\pm18$     &  $121\pm22$     &  $147\pm16$     &  $140\pm16$     &  $132\pm15$     \\  
$k T\,{(\rm keV})_{mbb}$                   & $0.931\pm0.021$ & $0.864\pm0.025$ & $0.705\pm0.013$ & $0.691\pm0.013$ & $0.680\pm0.013$ \\  
$\chi^2/d.o.f$                             &     103/89      &     75/82      &     84/80      &     108/81      &     95/81           \\
\hline                                                                                                                
\end{tabular}}

Notes: Fluxes and luminosities are estimated between 0.3--10\,keV. Fluxes are absorbed, whilst luminosities are unabsorbed.   \\
{The column density $N_{\rm H}$ of the neutral absorber and the temperature of the blackbody $k T_{bb}$
are coupled  between flux-resolved spectra extracted from the same observation. 
In fact, a preliminary individual fit  of each observation provided consistent results.}

\end{table*}
 
Previous work has suggested that the spectral shape of NGC\,55 changes
with the source flux: softer when fainter (Stobbart et al. 2004)
with the temperatures of the two blackbody components (tentatively)
found to be correlated (Pintore et al. 2015). 
Here we further investigate this problem using flux-resolved 
X-ray broadband spectroscopy.

\begin{figure}
  \includegraphics[width=1.3\columnwidth, angle=-90, bb=64 125 571 505]{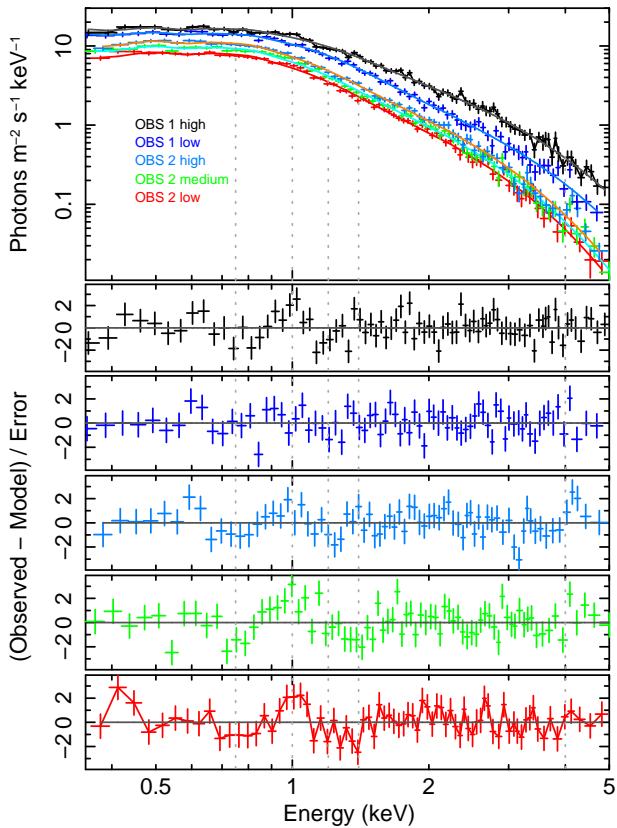}
   \caption{NGC 55 ULX EPIC flux resolved spectra with a continuum model 
         of blackbody plus modified disk blackbody 
         (see Table\,\ref{table:epic_flux_resolved_spectra}).
         The dotted lines show that the residuals could vary.} 
            \label{Fig:epic_flux_resolved_spectra}
\end{figure}

We split the two observations into five flux regimes.
A detailed description is provided in Appendix\,\ref{sec:epic_flux_resolved_lighcurves}.
Briefly, we extracted EPIC-pn lightcurves in the 0.3--10\,keV energy band for the two observations,
then chose two flux ranges for the first (shorter) observation and three flux ranges 
for the second observation, and then calculated the good time interval for each flux range.
The flux ranges were chosen such that the five spectra extracted in these 
good time intervals have comparable statistics.
The five flux-selected spectra are shown in Fig.\,\ref{Fig:epic_flux_resolved_spectra}.
A prominent knee appears at 1\,keV similar to that seen in some
ultraluminous supersoft X-ray sources 
(e.g. M\,101 and NGC\,247, Urquhart and Soria 2016).

In a preliminary fit performed to each of the five spectra with the best-fit 
\textit{bb}+\textit{mbb} continuum model (Sect.\,\ref{sec:epic_continuum})
we found that the column density $N_{\rm H}$
of the neutral absorber and the temperature of the cooler blackbody $T_{bb}$
did not significantly change between flux-resolved spectra selected from the same observation.
We therefore fit simultaneously the spectra of observation 1
coupling the $N_{\rm H}$ and the $T_{bb}$. 
The same coupling was repeated afterwards
when simultaneously fitting the spectra of the observation 2.

The flux-resolved spectral fits are shown in Fig.\,\ref{Fig:epic_flux_resolved_spectra}.
In Table\,\ref{table:epic_flux_resolved_spectra} 
we report the detail of the spectral fits
including the observed unabsorbed luminosities. 
The column density decreases with the increasing luminosity,
whilst the temperatures of both the cooler blackbody 
and the hotter multicolour blackbody increase.
The flux-resolved spectra confirm the residuals 
detected with a higher confidence
in the deeper, stacked, spectrum of the observation 2. 
Moreover, there is some tentative evidence of variability in the residuals
{(see Appendix\,\ref{sec:epic_gaussian_fits})}.
Deeper observations are necessary to study their variability in detail.

\section{Timing analysis}
\label{sec:timing}

NGC 55 ULX has shown an interesting temporal evolution in the last decade.
Stobbart et al. (2004) found evidence of dips in the lightcurve, 
which seem to have significantly smoothed in more recent \textit{Chandra} 
and \textit{Swift} observations (Pintore et al. 2015).
We therefore looked for short and long term variations.
We extracted the source and background lightcurves in the same regions
used to extract spectra (see Sect.\,\ref{sec:data})
and obtained a source background-subtracted lightcurve 
with the \textit{epiclccorr} task.

\begin{figure*}
  \includegraphics[width=1.0\columnwidth]{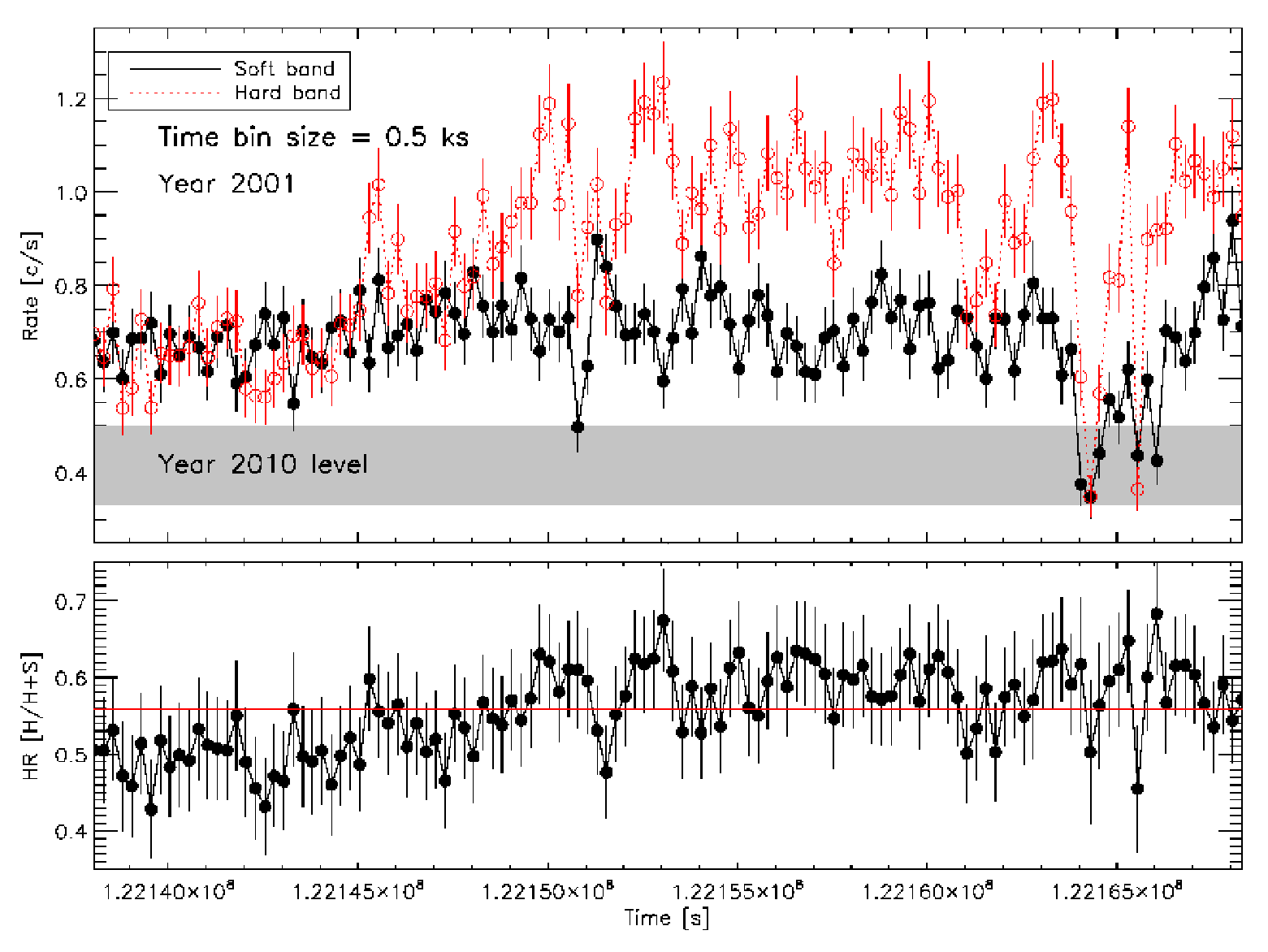}
  \includegraphics[width=1.0\columnwidth]{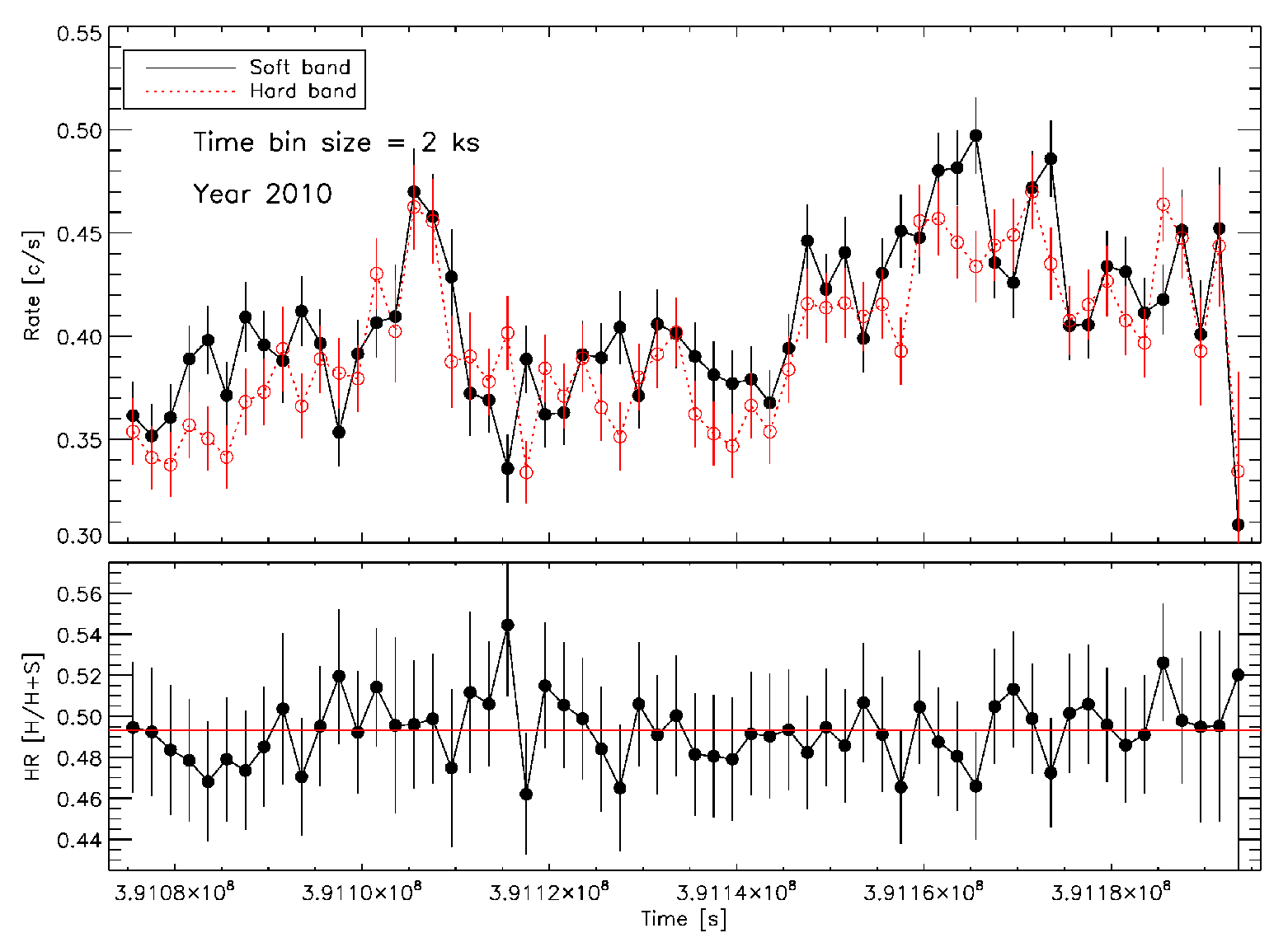}
   \caption{EPIC/pn lightcurves of the first (left) and second (right) observations,
            The sizes of the time bins are different between the exposure (500\,s and 2\,ks)
            due to the different fluxes. The soft band refers to the 0.5--1\,keV
            energy range and the hard band refers to 1--5\,keV.
            For more detail on the energy band selection see Sect.\,\ref{sec:lightcurves}.
            Notice the lower flux and variability in 2010.
            The area shaded in grey in the left panel includes
            the count rate throughout the whole 2010 observation
            which seems to match the dips seen in 2001.} 
            \label{Fig:lightcurves}
\end{figure*}

\begin{figure}
  \includegraphics[width=0.95\columnwidth]{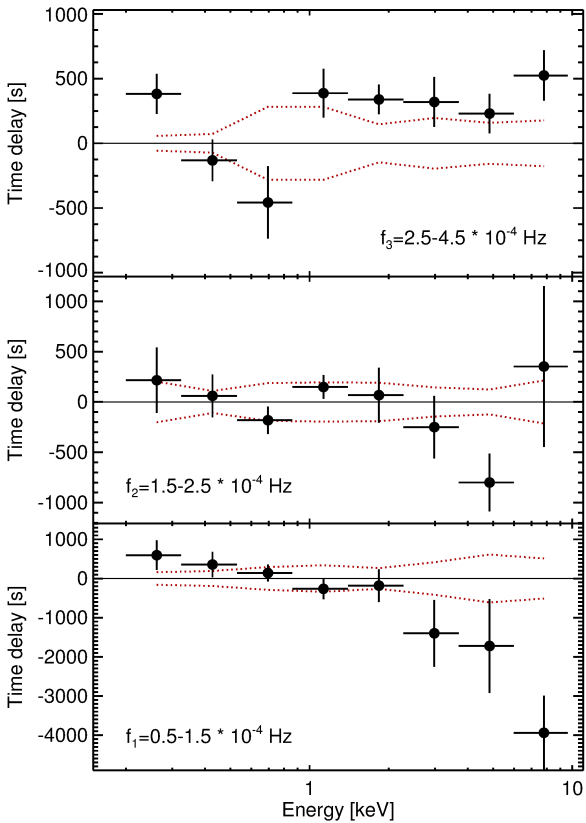}
   \caption{Lag-energy spectra for different frequency ranges 
            with the NGC 55 ULX EPIC-pn (observation 1). 
            Note the trend towards lower energies at low frequencies
            in the bottom panel.
            {The dotted red lines show the magnitude expected from the 
            Poisson noise. The points 
            should be randomly distributed between these lines 
            in absence of any lag.}} 
            \label{Fig:lags}
\end{figure}

\begin{figure}
  \includegraphics[width=0.95\columnwidth]{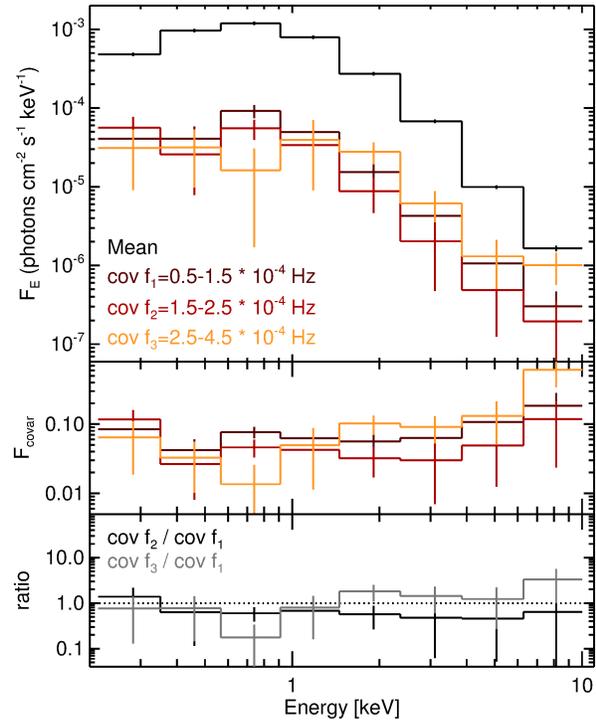}
   \caption{Covariance spectra for the same frequency ranges 
            as Fig.\,\ref{Fig:lags} (observation 1). 
            The covariance spectra follow the shapes of the lags 
            measured at the corresponding frequency. 
            The middle panel shows the fractional variability calculated in 
            each frequency range and divided by the mean rate.
            The bottom panel shows the ratios between 
            the fractional variability at  
            some frequencies.} 
            \label{Fig:covariance}
\end{figure}
       
\subsection{Lightcurves}
\label{sec:lightcurves}

NGC 55 ULX lightcurves are shown in Fig.\,\ref{Fig:lightcurves}. The lightcurves
were extracted in a soft (0.5-1.0\,keV) and a hard (1--5\,keV) energy band
in order to study the behaviour of the source on the low-energy 
and high-energy side, respectively, of the 1\,keV bump seen in its EPIC spectrum
as well as in the spectra of several ultraluminous X-ray sources
(see e.g. Pintore et al. 2015, Urquhart and Soria 2016).

The source behaviour has dramatically changed in the last decade.
During the observation of 2001 the hard band was highly variable, 
exhibiting strong dips as previously shown by Stobbart et al. (2004).
During the observation of 2010 the source was much less variable, 
with the hard and soft band showing the same variability pattern
(see Fig.\,\ref{Fig:lightcurves}). Interestingly, 
the count rate in 2010 is consistent with the dips observed during 
the first 2001 observation. 

{We investigated the variability amplitude in each observation 
as a function of temporal (Fourier) frequency using power density spectra.  
We follow the standard method of computing a periodogram in 10 ks lightcurve 
segments before averaging over the segments at each frequency bin, 
and subsequent binning up over adjacent frequency bins 
(van der Klis 1989, Vaughan et al. 2003). 
The resulting power density spectra are shown 
in Fig.\,\ref{Fig:pds}} and confirm that the source 
has very limited variability in 2010 compared to 2001 (see also Heil et al. 2009). 
We therefore focus on the 2001 observation only in the remainder of this Section. 

\subsection{Time lags and covariance spectra}
\label{sec:lags}

We studied the timing properties of NGC 55 by calculating the cross-spectrum between 
light curves in different energy bands. A detailed description of the method 
can be found in Alston, Vaughan \& Uttley (2013) and Uttley et al. (2014, and references therein).  
We calculated the cross-spectrum at each Fourier frequency in 10 ks segments and then averaged
over the segments at each frequency, before binning in geometrically spaced frequency 
bins by a factor of 1.6.  In a given frequency band we calculated the time lags between
a comparison energy band versus a broad (in energy) reference band, 
whilst subtracting the comparison band from the reference band if it fell within it.  
In this way, the time lag in each energy bin is the average lag or lead of the band 
versus the reference band. 
We used the $0.3 - 1.0$\,keV band as reference where the rms is high. 
In Fig.\,\ref{Fig:lags} we show the resulting 
lag-energy spectrum for three low frequencies. 

The power density spectrum is very flat and the Poisson noise dominates above a mHz 
(see, e.g., Heil et al. 2009).  
We selected three frequency intervals for which it is possible to measure any lag and 
covariance. Above these frequencies the lags are consistent with zero
and a lot of the covariance energy bins are negative as the noise dominates. 
{We also estimated the lag magnitude expected from the Poisson noise 
contribution to the phase difference using Eq.\,30 of Vaughan et al 2003,
see red dotted lines in Fig.\,\ref{Fig:lags}.
In the absence of any intrinsic lag we should expect the bins to be randomly 
distributed between these limits, which is not the case at least
at high and low frequencies.}

At low frequencies ($0.5-1.5\times10^{-4}$\,Hz in Fig.\,\ref{Fig:lags} bottom panel) 
there is evidence for the softer bands lagging behind harder bands. 
As the frequency increases (from bottom to top panel), 
there is a systematic change in the lag above 0.9 keV
with the profile flattening and possibly
the hard bands lagging the soft at higher frequencies ($2.5-4.5\times10^{-4}$\,Hz).  
There is a tentative a sharp change in the lag below $0.9$\,keV.

To better understand the behaviour in the lag energy spectra around 1 keV, 
we computed the covariance spectra at each Fourier frequency
(Wilkinson \& Uttley 2009).  
The covariance spectra show the correlated variability between two energy bands, 
or between a comparison band and a broad reference band (minus the comparison band).  
We computed the covariance from the cross-spectrum following Uttley et al. (2014).

The covariance spectra for the same three frequencies are shown 
in Fig.\,\ref{Fig:covariance} (top).
The overall (mean) covariance spectrum mimics the shape
of the energy spectrum (see Fig.\,\ref{Fig:epic_flux_resolved_spectra}.
The covariance spectra extracted in different frequency ranges
follow the trend of the energy lag spectra (see Fig.\,\ref{Fig:lags}).
The covariance increases towards lower energies, but then flattens 
or possibly drops below 1\,keV. 
On average, the high-frequency covariance spectrum
appears to be harder than the low-frequency one,
with a possible drop below 1\,keV,
suggesting the presence of different variability processes 
(Fig.\,\ref{Fig:covariance} middle panel).
{The covariance spectra extracted at different frequencies
are very similar with the current statistics,
but more data would certainly help to distinguish some differences
at low and high frequencies. The same applies to the lag-energy spectra.}
More work will follow up in a forthcoming paper to provide better 
constraints on these trends and to compare the results 
with other ULXs.

\section{Discussion}
\label{sec:discussion}

Urquhart and Soria (2016) modelled the soft X-ray spectral features 
and the $\sim$\,1\,keV ($\sim12$\,{\AA}) drop in several CCD spectra of ULSs with a model 
of thermal emission and an absorption edge. They interpreted the features
as a result of absorption and photon reprocessing by an optically-thick wind
which obscures the innermost regions where most hard X-rays are produced.
This suggested that ULSs are actually ULXs viewed almost edge on
or maybe their outflow is thicker, at the same viewing angle. 
A denser, more optically thick outflow may be the result of higher accretion rates 
above Eddington, assuming that \.m$_{\rm wind}$ is proportional to \.m$_{\rm disk}$
(e.g., Shakura and Sunyaev 1973). 
Moreover, at higher super-Eddington rates, the outflow launching region moves to larger radii 
on the disk, where the escape velocity and the Compton temperature are lower, 
which favours denser winds (e.g., Poutanen et al. 2007).
Motivated by the recent detection of relativistic winds in the NGC\,1313 and NGC\,5408 
ULX-1 high-resolution RGS spectra taken with XMM-\textit{Newton} (see Paper\,I), 
we searched for a deep RGS observation of an object that shows the properties 
of both the ULX and ULS states and found NGC 55 ULX. 

\subsection{Discovery of a powerful wind}
\label{sec:discussion_wind}

The primary goal of our study was the search for unambiguous signatures of a wind.
The detection of (unresolved) spectral residuals in previous work done on NGC 55 ULX 
was certainly encouraging (e.g. Middleton et al. 2015b).
We confirmed the presence of these residuals particularly in the 2010
EPIC deep spectrum (see Figs.\,\ref{Fig:epic_continuum} and \ref{Fig:epic_flux_resolved_spectra}).

The evidence hinting at a wind was confirmed when
we looked at the high-resolution RGS spectrum of NGC 55 ULX,
which shows a wealth of emission and absorption features located at energies that 
are blueshifted compared to the energies of the strongest atomic transitions 
in the soft X-ray energy band (see Figs.\,\ref{Fig:rgs_spectrum} and \ref{Fig:line_detection}).
Accurate spectral modelling shows the presence of an outflowing plasma
with a complex ionization and velocity structure 
(see Figs.\,\ref{Fig:rgs_spectrum_fit} and \ref{Fig:wind_structure_IDL}).
The best-fitting models suggest photoionization, 
but we cannot exclude some contribution 
from collisional ionization, particularly for high-ionization species.
The gas responsible for the emission lines moves with velocities $(0.01-0.08)c$
lower than those exhibited by the absorbing gas $(0.06-0.20)c$,
see Table\,\ref{table:improvements} and Fig.\,\ref{Fig:wind_structure_IDL}.
Both the emission and the absorption lines indicate that the ionization state 
increases with the outflow velocity suggesting that 
we are detecting hotter/faster phases coming from inner regions.
Our limited statistics prevents us from claiming the detection of more than 
two photoionized emitters and absorbers in the line of sight, but additional
observations would certainly help to understand the ionization and velocity
structure of the wind in NGC 55 ULX.

An interesting result is the spectral signature of the high-ionization 
($\log \xi \sim 3.3$) absorber with the largest column density.
This component absorbs a substantial amount of photons 
below 12\,{\AA} (or above 1\,keV),
which provides a natural explanation for the spectral shape of NGC 55 ULX 
and the drop seen in several ULSs (e.g. Urquhart and Soria 2016). 
The main difference is that such a turnover is more pronounced in ULSs because they
are seen more edge on or because the photosphere of the wind is simply further out 
due to a larger accretion rate (see e.g. Middleton et al. 2015a 
and Urquhart and Soria 2016) 
with resulting increased obscuration of the inner regions.

{It is useful to compare the wind power to the bolometric radiative luminosity
of the source. The kinetic luminosity of the outflow or wind power $L_{\rm kin} = 1/2 \dot{M} v^2_{\rm out}$
can be expressed as $L_{\rm kin} = 2 \pi L_{\rm ion}\,m_{\rm p}\,\mu\,v^3_{\rm out}\,C_V\,\Omega / \xi$,
where $L_{\rm ion}$ is the ionizing luminosity, $m_{\rm p}$ is the proton mass, 
$\mu$ is the mean atomic weight ($\sim1.2$ for solar abundances), 
$v_{\rm out}$ is the outflow velocity, $C_V$ is the volume filling factor (or `clumpiness'), 
$\Omega$ is the solid angle, and $\xi$ is the ionization parameter of the wind.
We calculated the ratio between the wind power and the bolometric luminosity for the
fast and most significant absorber (see Table\,\ref{table:improvements}) and obtained:
$L_{\rm kin} / L_{\rm bol} = 1300 \pm 650\,C_V\,\Omega\,L_{\rm ion}/L_{\rm bol}$.
According to the simulations of Takeuchi et al. (2013), a wind driven by a strong radiation
field in super-Eddington accretion has a typical clumpiness factor of $\sim0.3$
and is launched over wide angles of $10^{\circ}-50^{\circ}$ from the disk rotation axis. 
A significant fraction of the source radiative luminosity ($L_{\rm bol}$) must have been
used to ionize the wind before some reprocessing occurred.
In our broadband EPIC fits we found that, on average, the 0.3--10\,keV total source flux
was equally partioned between the hard (disk) and soft (reprocessed) components.
Therefore, it is reasonable to adopt $L_{\rm ion}/L_{\rm bol}\sim0.5$.
This provides $L_{\rm kin} / L_{\rm bol} \sim 30-100$,
which is similar to that measured for NGC 1313 X-1 and NGC 5408 X-1 (see paper\,I and 
Walton at al. 2016a), but is an extreme value 
if compared to the strongest outflows seen from sub-Eddington systems, e.g. active
galactic nuclei (e.g., Fabian 2012, and references therein).}

\subsection{Long and short term spectral variability}
\label{sec:discussion_variability}

We studied the timing and spectral properties of NGC 55 ULX to understand
if {they also suggest that the source is a transition between {classical}} ULXs and ULSs.

If the thick, outflowing, wind scenario is correct then we should expect a correlation 
between the spectral hardness of the source and the X-ray flux (e.g. within 0.3--10\,keV).
For instance, if either the accretion rate or the inclination decrease
then the optical depth decreases, which corresponds to a lower down-scattering
of the hard photons produced in the inner regions.
We test this picture by splitting the observations in different flux regimes
which provided five high-quality EPIC spectra (see Fig.\,\ref{Fig:epic_flux_resolved_spectra}).
We modelled the EPIC spectra with a spectral continuum consisting of a blackbody
and a multicolour blackbody emission components in agreement with Pintore et al. (2015)
and with our search for the best-fitting continuum.
The temperatures of both components (particularly the hotter multicolour blackbody)
significantly increase with the luminosity supporting the wind scenario
(see Table\,\ref{table:epic_flux_resolved_spectra}).
This is confirmed by the observed decrease in column density of the neutral absorber
with luminosity (the Galactic interstellar medium account less than 10\,\%
according to the H\,I maps and does not affect our result).
In support for this scenario we also found that the emitting area of the 
soft ($\sim0.2$\,keV) blackbody component increased by a factor 2 from the brighter (2001) 
to the fainter (2010) observation in agreement with the study of ULSs
of Urquhart and Soria (2016) and Feng et al. 2016,
where the expansion of the photosphere and the decrease of the temperature
shift the peak of the spectrum from the X-ray to the far-UV energy band.
According to our spectral fits, the radius of the photosphere is 2000--3000\,km 
and the temperature well above 0.15\,keV, which explains why
NGC 55 still looks like ULX rather than a ULS 
according to the classifications of Sutton et al. (2013) and Urquhart and Soria (2016).
This may suggest that either the inclination or the accretion rate
in NGC 55 is smaller than in typical ULSs.


There are no striking dips in the 2010 exposure if compared to those in 2001,
which is not surprising since they were not detected in most recent observations 
taken with \textit{Chandra} and \textit{Swift} (see Pintore et al. 2015).
Lightcurves extracted with the same binning are shown 
in Appendix \ref{sec:epic_flux_resolved_lighcurves}.
We remark that the flux level in 2010 is consistent 
with the lowest level of 2001, suggesting that the source
is in a semi-obscured state (see Fig.\,\ref{Fig:lightcurves}). 
The lack of variability and the low flux in 2010 is consistent with a picture where 
the innermost regions, which produce the high energy hard X-rays 
and the high variability, are partly obscured by an intervening absorber
in agreement with the hypothesis of Stobbart et al. (2004),
Middleton et al. (2011) and Sutton et al. (2013). 
Either the accretion {rate increased, further inflating the outer regions
of the disk,} or the line of sight changed (e.g. due to precession)
with further obscuration of the innermost ``hotter'' regions.

The tentative detection of a large soft lag, $\sim$\,1000s, at low frequencies 
further strengthens our scenario (see Fig.\,\ref{Fig:lags})
suggesting that the soft emitting gas in the wind and the upper disk
photosphere are reprocessing the photons produced in the innermost regions.
The covariance spectra computed within different frequency ranges
follow the shape of the lag-energy spectra (see Fig.\,\ref{Fig:covariance}),
confirm this picture and show evidence for 
two possible main processes that produce variability:
fluctuation within the inner disk (harder) and 
slow reprocessing through the wind (softer).
{It is also possible that when the column density of the wind decreases
the wind becomes optically thin, but the hard X-ray photons start penetrating the wind 
before the soft photons (see, e.g., Kara et al. 2015). 
This may explain the source hardening when 
the luminosity increases. 
In a future project we will investigate all possible scenarios 
by comparing more sources and lags at different time scales.}


Our results agree with Middleton et al. (2015a). 
They computed the covariance spectra on a broad frequency band 
and showed a significant lack of variability in the 2010 observation,
suggesting that the variable, hard, X-ray component 
has intercepted cooler/optically thicker material.
They also found that the shape of the covariance spectra
in NGC 55 ULX and several other ULXs was consistent with the variability originating 
in the hard component only, which agrees with a model where the variability on short 
and long timescales at moderate inclinations is dominated by obscuration 
of the high energy emission.

The lags tentatively detected here appear at much lower frequencies than those
shown by Heil et al. (2010) and De Marco et al. (2013) in NGC 5408 X-1. 
Hernandez-Garcia et al. (2015) also found evidence of a $\sim$\,1\,ks
time lag in NGC 5408 X-1 and showed that the time delays are energy-dependent
and that their origin is not related 
to reflection from an accretion disk (`reverberation').
They also argued that associating the soft lag with a quasi-periodic
oscillation (QPO) in these ULXs, drawing an analogy between soft lags in ULXs 
and soft lags seen in some low-frequency QPOs of Galactic X-ray binaries,
is premature.

The two orders of magnitude in time delays (from a few seconds
to hundred seconds in NGC 55 and NGC 5408 ULXs) point towards 
a complex energy lag spectrum and we also caution from using lags 
to derive mass estimates without a comprehensive analysis 
in a large frequency range.
Although this is beyond the scope of this paper, we provide
a simple argument to compare lags in AGN, X-ray binaries, and ULXs.
The lag magnitude in NGC 55 ULX as a function 
of the variability timescale is $\sim$10\% (1000s; 1/f=10,000s). 
The hard lags in AGN are typically 1\%, 
the AGN soft lags are $\sim$2\% (e.g., Alston et al. 2013), 
whilst in hard state X-ray binaries they are 0.5-1\% (e.g., Uttley et al. 2011).
De Marco et al. (2013) measured about $\sim$5\%  
for the higher frequency lags in NGC 5408 X-1. 
Hence, it is possible that
the processes causing these lags are different. 
The long soft lags in NGC 55 ULX and NGC 5408 X-1
{could} be related to some phenomena occurring 
in the outer region.
In fact, the 1000s magnitude is likely due to the combination of
light travel time plus thermalization in the wind and 
additional local scattering processes before 
the photons are re-emitted towards the observer.
Such phenomena imply a distance that can be large enough to damp the variability
with a corresponding decreasing of the covariance at high frequencies
(see Fig.\,\ref{Fig:covariance}).
The short (a few seconds) soft lags seen by 
Heil et al. (2010) and De Marco et al. (2013) can be different 
processes occurring in the inner regions. 

\subsection{NGC\,55 vs ULXs \& ULSs: the big picture}
\label{sec:discussion_comparison}

A lot of work has already been done in order to classify ULXs and ULSs 
and to understand the effect of the inclination (viewing angle) and the 
accretion rate on their spectral and timing behaviour
(see, e.g, Gladstone et al. 2009, Sutton et al. 2013, 
Middleton et al. 2015, Urquhart and Soria 2016, and references therein).
Therefore, we briefly {highlight why} NGC 55 ULX looks like a transitional 
form between ULXs and ULSs for the shape of its spectral continuum.
More attention will be given on the way the detection of winds
fits in the framework of super-Eddington accretion.

\subsubsection{The spectral shape}
\label{sec:discussion_continuum}

The higher-energy broad-band component is an interesting link 
between ULXs and ULSs.
The sequence from hard ULXs to soft ULXs and then to ULSs
shows a progressively lower temperature 
(whether modelled as Comptonization or a modified disk,
e.g., Gladstone et al. 2009). 
The temperature is therefore a function of \.m$_{\rm wind}$ 
and/or the viewing angle. 
The optical depth $\tau$ (for a Comptonization model) instead increases 
along the same sequence, from $\tau \sim$ few for hard ULXs 
to $\tau \gtrsim$ 10 for soft ULXs, 
to $\tau$ $\longrightarrow$ infinity for ULSs.
Alternatively, the photon index $\Gamma$ increases from 
$\sim$ 1.5 to $\sim$ 2.5 and then $>$ 4 along the same sequence
in analogy with the ($T$,$\tau$) changes for a comptonizing region.  
The hard component of NGC 55 can be well modelled as 
either a steep ($\Gamma>4$) power law or as a modified disk, 
which essentially means a large $\tau$ 
(see e.g. Tables \ref{table:continuum} and 
\ref{table:epic_flux_resolved_spectra}). 
{We have also shown that the shape of NGC 55 ULX spectrum matches 
that of the high-flux state of NGC 247 ULX, an object that
shifts between the ULX and the ULS classical states (see Fig.\,\ref{Fig:ULX_sequence}).}
This suggests that NGC 55 ULX is indeed a transitional object
between the ULXs and ULSs. 

\subsubsection{The spectral features}

The emission and absorption features detected in NGC 55 ULX are produced by the same
ionic species detected in NGC 1313 X-1 and NGC 5408 X-1 RGS spectra (see Paper\,I).
For instance, the properties of the most significant absorber detected in NGC 55 ULX 
(see e.g. Table\,\ref{table:improvements}) seem to match those of the fast ($0.2c$) 
component in the other two ULXs. 
The detection of a multiphase structure in NGC 55 ULX would point towards 
a tighter analogy to NGC 5408 X-1, which also exhibits a wind more complex 
than NGC 1313 X-1, with different ionization states and outflow velocities.
We notice that evidence for an edge-like feature around 11\,{\AA} 
was also found in the RGS spectrum of NGC\,6946 X-1 (see Paper\,I), 
which sometimes exhibits very soft spectra 
like NGC 55 and NGC 5408 (the soft ultraluminous regime, Sutton et al. 2013).
Unfortunately the interpretation of this feature in NGC 6946 X-1 was difficult due to
its flux at 1 keV (OBS ID 0691570101) $\sim50$\% lower than NGC\,55 ULX.

\begin{figure}
  \includegraphics[width=0.95\columnwidth]{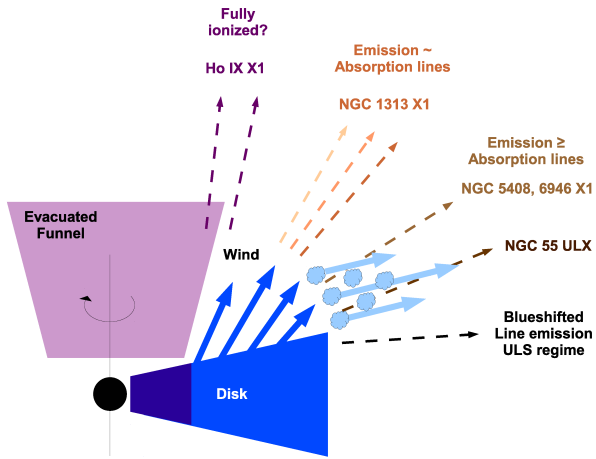}
   \caption{Simplified scheme of high mass
   accretion rate sources (see also Middleton et al. 2011). 
   The light blue region shows the soft X-ray emission of the accretion disk, 
   altered by a photosphere of a radiatively-driven optically-thick
   wind. The dark blue region, closer to the compact
   object is dominated by highly variable, optically-thinner,
   turbulent Comptonization emitting high-energy ($>1$ keV) X-rays.
   The dashed lines indicated some possible sightlines for some famous ULXs
   and for the ULS regime (see e.g. Urquhart and Soria 2016).
   We believe that the line of sight of NGC 55 ULX is somewhere between 
   the classical ULX and ULS sources, but still at high inclination 
   where outflowing material is both emitting and absorbing
   photons from the inner regions.} 
            \label{Fig:geometry}
\end{figure}

The presence of \textit{blueshifted} emission lines in NGC 55 ULX suggests that 
we may be looking through a different line of sight compared to NGC 1313 and 5408 ULXs. 
The continuum is for instance significantly lower than NGC 5408 X-1 
(see Extended Data Fig.\,4 in Paper\,I), 
but at 1\,keV NGC 55 ULX is comparable if not brighter,
with the emission lines at other energies also 
matching the flux of the lines in the other ULXs.
Remarkably, the line-emitting gas phases account for 10--20\,\% of the total source flux 
in the X-ray (0.2--10\,keV) energy band in agreement with NGC\,5408 X-1,
while in NGC 1313 X-1 the lines contribute about 5\,\% of the source flux.
If the emission lines in these ULXs have the same origin then we should
expect that NGC 55 ULX is seen through a preferential line of sight
where the outflow velocity of the emission component is almost maximum
(see Fig.\,\ref{Fig:geometry}),
while the other three ULXs may be seen through a line of sight 
where the motion of the line-emitting gas is more tangential.
{This would explain why the emission lines 
in NGC 5408 X-1 appear to be broader by an order of magnitude 
($\sigma_v\sim1000$\,km s$^{-1}$, see Paper\,I).}

NGC 1313 X-1 shows a large range of spectral hardness with strong absorption 
features at $\sim0.2c$, which suggests moderate inclination angles with the 
wind variability possibly due to precession (see Middleton et al. 2015b).
Holmberg IX X-1 shows very shallow residuals in EPIC spectra 
and a spectral hardness among the highest measured in ULXs
which argues in favour of a nearly face on view 
(e.g. Sutton et al. 2013, Middleton et al. 2015a, {Luangtip et al. 2016}, Walton et al. 2016c).

{The emission features detected in NGC 55 ULX are produced by the same ionic species
that give origin to the blueshifted emission detected in the relativistic jet of the 
Galactic super-Eddington accretor SS 433 (see, e.g., Marshall et al. 2002 and references therein).
Similar, powerful, outflows have been discovered in Galactic black holes X-ray binaries
such as IGR J17091--3624 (King et al. 2012), MAXI J1305--705 (Miller et al. 2014),
4U 1630--47 (Diaz-Trigo et al. 2013) and during superburst of the neutron star 
X-ray binary SAX J1808.4--3658 (Pinto et al. 2014).}

\subsubsection{Ultraluminous supersoft sources}

A comprehensive analysis of the ULX phenomenology also needs 
high-resolution spectroscopy of ULSs in order 
to study the cases of extreme absorption and to ultimately confirm an 
unification scenario for ULXs and ULSs 
(see Fig.\,\ref{Fig:geometry} and, e.g., Poutanen et al. 2007, Urquhart and Soria 2016,
and references therein).
The ULS archive has XMM-\textit{Newton} exposures that are just too short
($\lesssim50$\,ks) to provide RGS spectra with statistics good enough
to significantly detect sharp spectral features.
NGC 247 ULX is likely the best candidate among all ULSs because of its brightness,
good isolation, and sharp drop at 1\,keV (see e.g. Urquhart and Soria 2016, {Feng et al. 2016}).
The RGS archive has two on-axis exposures with a modest 30\,ks exposure time each.
These prevent us from doing an accurate analysis, 
{particularly for the low-flux ULS spectrum,}
which would require more data
or additional spectra of other ULSs 
and, therefore, we defer this work to {future work}.

\begin{figure}
  \includegraphics[width=0.95\columnwidth, bb=70 50 535 496, angle=-90]{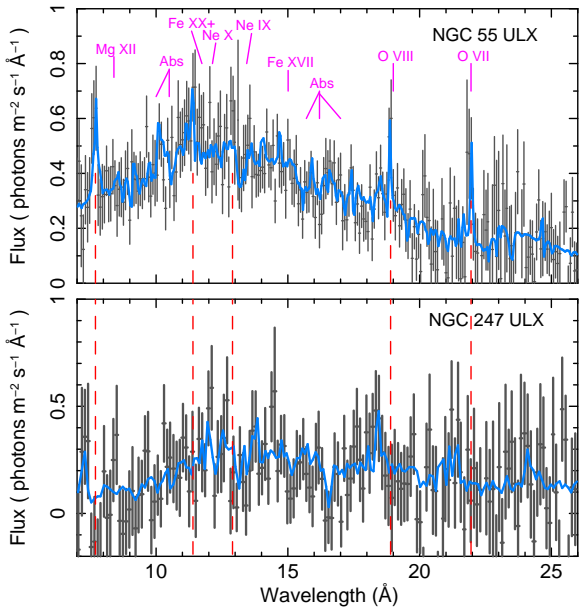}
   \caption{NGC 55 ULX RGS spectrum with overlaid the structured with model (top)
         compared to NGC 247 ULX with a similar wind 
         model on top of a blackbody emission continuum (bottom):
         low-$\xi$ slow ($v=0.05c$) and high-$\xi$ fast ($v=0.13c$)
         photoionized emission components and 
         high-$\xi$ fast ($v=0.14c$) photoionized absorption.
         The red-dashed lines mark some relevant blueshifted emission line detected in NGC 55 ULX.
         Some interesting features appear on their blue side in NGC 247 ULX.} 
            \label{Fig:rgs_spectrum_fit_ngc247}
\end{figure}

{Here we present a preliminary result.
Briefly, we repeated the RGS analysis done in this work on NGC 55 ULX 
with the {high-flux} RGS exposure of NGC 247 ULX
(ID 0728190101, see Sect.\,\ref{sec:source}).
The spectrum also shows hints for absorption and emission features,
fainter than in NGC 55 ULX due to the worse statistics
(see Fig.\,\ref{Fig:rgs_spectrum_fit_ngc247}). 
In particular, RGS confirms the spectral bending below 12\,{\AA}
(above 1\,keV) as seen in CCD spectra (see Fig.\,\ref{Fig:ULX_sequence}).
It is encouraging that the spectral features can be modelled 
with a wind model similar to that used in NGC 55 ULX, 
i.e. two photoionized emitters and one photoionized absorber,
with the ionization increasing with the outflow velocity
(see Fig.\,\ref{Fig:rgs_spectrum_fit_ngc247}). 
The velocity of the line-emitters in NGC 247 ULX seems
to be even higher than those in NGC 55 ULX showing a further preferential 
location of the line of sight.}
{Jin et al. (2011) have analyzed the ULS-state EPIC-pn spectrum
of NGC 247 that we have shown in Fig. 1 and found similar residuals 
to those detected in NGC 55 ULX. In particular, 
absorption features at 0.7 and above 1 keV as well as emission at 
or just below 1 keV. As suggested by Feng et al. (2016) and Urquhart and Soria (2016),
the supersoft ultraluminous (SSUL or ULS) regime is likely an extension 
of the soft ultraluminous state toward higher 
high accretion rates with the blackbody emission arising from the 
photosphere of thick outflows and the hard X-rays being emission 
leaked from the embedded accretion disk via the central low-density 
funnel or advected through the wind. 
The wind thickens throughout the ULX-SUL-SSUL sequence and imprints
the spectral curvature and the absorption edge at 1 keV.}

\subsection{XMM-Newton in the next decade}
\label{sec:next_decade}

Ideally, we would need a statistical sample to probe the scenario shown 
in Fig.\,\ref{Fig:geometry} comparing the properties of the wind in sources  
at different angles of inclination to the observer and mass accretion rate.
Only a handful of sources have been observed with deep, on-axis, XMM-\textit{Newton} 
observations with statistics good enough to enable RGS analysis
and no on-axis full-orbit observation of any bright ULS has ever been taken.
The fact that the wind lines can be studied imply that with a long exposure of a ULS 
we will learn more about their physics.
Deeper observations of other ULXs and ULSs will provide a complete sample 
of ultrafast winds, 
while deeper observations of those sources which already have well-exposed RGS spectra 
(e.g. NGC 1313 X-1 and NGC 55 ULX) will enable to study the wind in detail 
and its dependence on flux, spectral hardness, 
and source quasi-periodicities (e.g. Walton et al. 2016b). 
All this is crucial to understand the relation between the wind and both the accretion rate 
and the viewing angle, which is useful to understand the launching mechanism.

This research field is still rather new and the ESA's XMM-\textit{Newton} satellite
has unique capabilities to further develop it. 
The gratings on board NASA's \textit{Chandra} satellite (HETGS/LETGS,
e.g. Brinkman et al. 2000 and Canizares et al. 2005) have the highest spectral
resolution available in the X-ray band, but their effective area is smaller 
than that of RGS in the soft X-ray band where most ULX wind features are shown
and where the flux is highest.
HETGS and LETGS are clearly optimal to distinguish spectral features in 
the brightest objects, whilst
the RGS has unique capabilities to detect lines in weak objects,
which means that the continuation of the XMM-\textit{Newton} and Chandra missions 
is crucial to study the ULXs and ULSs in detail.

{ULXs and ULSs will be excellent targets for a $Hitomi$ replacement mission,
for the proposed $Arcus$ gratings mission 
and, particularly, for the two-square-meter ESA's $Athena$ mission.
$Arcus$ can significantly improve the RGS results due its much higher 
spectral resolution in the soft X-ray energy band (e.g., Kaastra 2016).
$Hitomi$ has already proven that we can change the way we perform X-ray spectroscopy
thanks to its combination of excellent spectral resolution and collective
area in the high energy X-ray band (Hitomi Collaboration 2016).
$Athena$ will completely revolutionize X-ray astronomy
due to its combination of high spatial and spectral resolution
(Nandra et al. 2013).}

We remark that the study of the accretion flow in ULXs and ULSs is important
to understand the phenomenology of super-Eddington accretion, 
which may be required to occur during the early stages of the Universe
to build up the supermassive black holes that have been discovered to power AGN at high redshifts
(see e.g. Volonteri et al. 2013).

\section{Conclusions}
\label{sec:conclusion}

In the last decade it has been proposed that a substantial fraction
of the population of ultraluminous X-ray sources (ULXs) and 
ultraluminous supersoft sources (ULSs) are powered by super-Eddington
accretion onto compact objects such as neutron stars and 
black holes (see, e.g., King et al. 2001, Roberts 2007,
Urquhart and Soria 2016, Feng et al. 2016, and references therein). 
In particular, ULSs could be a category of ULXs 
observed at high inclination angles, possibly edge on,
where a thick layer of material is obscuring the innermost hard X-ray 
emitting regions (e.g., Kylafis \& Xilouris 1993 and Poutanen et al. 2007). 
We have studied the NGC 55 ULX which we believe is a hybrid source
showing properties common to both ULX and ULS.
The presence of a spectral curvature at 1\,keV and a high energy tail
that can be described as a $\gtrsim0.7$\,keV disk-like blackbody 
place it just in between these two categories of X-ray sources.

We have found a powerful wind characterized by emission and absorption 
lines blueshifted by significant fractions of the speed of light
($0.01-0.20c$) in the XMM-\textit{Newton}/RGS spectrum of NGC 55 ULX.  
The detection of such a wind is consistent with the predictions
of super-Eddington accretion (Takeuchi et al. 2013). 
The wind has a complex dynamical structure with an ionization state
that increases with the outflow velocity, which indicates 
launching from different regions of the accretion disk.
The comparison of the wind in NGC 55 ULX with that detected in other ULXs
(Pinto et al. 2016)
suggests that the source is being observed at high inclinations,
but not high enough to look exactly like a ULS,
in agreement with the classification on the basis 
of its spectral shape (e.g., Sutton et al. 2013).
However, the strongest wind component partly absorbs the source flux below 1\,keV,
generating a drop similar to that observed in ULSs.

The long and short term spectral variability of the source 
shows a softening of the spectrum at lower luminosities,
i.e. around the Eddington luminosity 
{(for a $10\,M_{\odot}$ black hole)},
which agrees with the proposed scenario of wind clumps 
crossing the line of sight and partly obscuring the innermost
region where most hard X-rays come from
(e.g. Middleton et al. 2015a, and references therein).

We have found evidence for a long $\sim$1000s soft lag at low frequencies,
similar to NGC 5408 X-1 (Hernandez-Garcia et al. 2015),
which may indicate that part of the emission coming from the inner regions
has been reprocessed in the outer regions before being re-emitted
towards the observer. This provides further support to the wind scenario.

Deeper XMM-\textit{Newton} observations of NGC 55 ULX and  
other ULXs and ULSs will enable a detailed study
of the dependence of the wind on the accretion rate and the inclination angle,
which can help us to understand the geometry
and the accretion flow in these extraordinary astronomical objects 
and in super-Eddington accretors in general.

\section*{Acknowledgments}

This work is based on observations obtained with XMM-\textit{Newton}, an
ESA science mission funded by ESA Member States and USA (NASA).
We also acknowledge support from ERC Advanced Grant Feedback 340442.
HE acknowledge support from the STFC 
through studentship grant ST/K501979/1. 
TPR acknowledges funding from STFC as part of the consolidated grant ST/L00075X/1.
DJW and MJM acknowledge support from  STFC via an Ernest Rutherford advanced grant.
We acknowledge the anonymous referee for useful comments that improved the paper.


\begin{thebibliography}{}

\bibitem[Alston, Vaughan \& Uttley (2013)]{2013MNRAS.435.1511A} Alston, W. N., Vaughan \& S., Uttley, P. 2013, MNRAS, 435, 1511

\bibitem[Bachetti et al. (2014)]{2014Natur.514..202B} Bachetti, M., Harrison, F., Walton, D.J., et al. 2014, Nature, 514, 202




\bibitem[Brinkman et al. (2000)]{2000SPIE.4012...81B} Brinkman, B. C., Gunsing, T., Kaastra, J. S. et al. 2000, SPIE, 4012, 81

\bibitem[Canizares et al. (2005)]{2005PASP..117.1144C} Canizares, C. R., Davis, J. E., Dewey, D. et al. 2005, PASP, 117, 1144

\bibitem[De Marco et al. (2013)]{2013MNRAS.436.3782D} De Marco, B., Ponti, G., Miniutti, G., Belloni, T., Cappi et al. 2013, MNRAS, 436, 3782

\bibitem[den Herder et al. (2001)]{2001A&A...365L...7D} den Herder, J. W., Brinkman, A. C., {Kahn}, S.~M., {Branduardi-Raymont}, G. et al. 2001, A\&A, 365, 7

\bibitem[Diaz-Trigo et al. (2013)]{2013Natur.504..260D} {D{\'{\i}}az Trigo}, M., {Miller-Jones}, J.~C.~A., {Migliari}, S., 
	{Broderick}, J.~W. and {Tzioumis}, T. 2012, Nature, 504, 260

\bibitem[Di Stefano \& Kong (2004)]{2004ApJ...609..710D} Di Stefano, R. and Kong, A. K. H. 2004, ApJ, 609, 710

\bibitem[Di Stefano \& Kong (2003)]{2003ApJ...592..884D} Di Stefano, R. and Kong, A. K. H. 2003, ApJ, 592, 884

\bibitem[Elvis (2000)]{2000ApJ...545...63E} Elvis, M. 2000, ApJ, 545, 63

\bibitem[Feng et al. (2016)]{2016ApJ...831..117F} Feng, H., Tao, L., Kaaret, P., Grise, F. 2016, ApJ, 831, 117

\bibitem[Fabian (2012)]{2012ARA&A..50..455F} Fabian, A. C. 2012, ARA\&A, 50, 455

  
\bibitem[Farrell et al. (2009)]{2009Natur.460...73F} Farrell, S. A., Webb, N. A., Barret, D., Godet, O., \& Rodrigues, J. M. 2009, Nature, 460, 73

\bibitem[Fuerst et al. (2016)]{2016arXiv160907129F} Fuerst, F., Walton, D.J., Harrison, F. A., Stern, D., {Barret}, D., {Brightman}, M., {Fabian}, A.~C., et al. 2016, accepted for publication in ApJL

\bibitem[Gladstone et al. (2009)]{2009MNRAS.397.1836G} Gladstone, J. C., Roberts, T. P. \& Done, C. 2009, MNRAS, 397, 1836

\bibitem[Greene and Ho (2007)]{2007ApJ...656...84G} Greene, Jenny E. and Ho, Luis C. 2007, ApJ, 656, 84

\bibitem[Heil \& Vaughan (2010)]{2010MNRAS.405L..86H} Heil, L. M. \& Vaughan, S. 2010, MNRAS, 405, L86

\bibitem[Heil et al. (2009)]{2009MNRAS.397.1061H} Heil, L. M., Vaughan, S., Roberts, T. P. 2009, MNRAS, 397, 1061

\bibitem[Hernandez-Garcia et al. (2015)]{2015MNRAS.453.2877H} {Hern{\'a}ndez-Garc{\'{\i}}a}, L., Vaughan, S., Roberts, T. P. and Middleton, M. 2015, MNRAS, 453, 2877

\bibitem[Hitomi Collaboration (2016)]{2016Natur.535..117H} Hitomi Collaboration 2016, Nature, 535, 117  

\bibitem[Jin et al. (2011)]{2011ApJ...737...87J} Jin, J., Feng, H., Kaaret, P., Zhang, S.-N. 2011, ApJ, 737, 87


\bibitem[Kaastra and Bleeker (2016)]{2016A&A...587A.151K} {Kaastra}, J.~S. and {Bleeker}, J. A. M. 2016, A\&A, 587, 151

\bibitem[Kaastra (2016)]{2016xnnd.confE...3K} {Kaastra}, J.~S. 2016, in proceedings of the XMM-Newton: The Next Decade conference, ESAC, ArXiv=1611.05924

\bibitem[Kara et al. (2015)]{2015MNRAS.446..737K} Kara, E., Zoghbi, A., Marinucci, A., Walton, D. J., Fabian, A. C. et al. 2015, MNRAS, 446, 737

\bibitem[King et al. (2001)]{2001ApJ...552L.109K} King, A. R., Davies, M. B., Ward, M. J., Fabbiano, G., \& Elvis, M. 2001, ApJ, 552, L109

\bibitem[King et al. (2012)]{2012ApJ...746L..20K} {King}, A.~L., {Miller}, J.~M., {Raymond}, J., {Fabian}, A.~C.,
	{Reynolds}, C.~S. et al. 2012, ApJ, 746, L20

\bibitem[Kong \& Di Stefano (2003)]{2003ApJ...590L..13K} {Kong}, A.~K.~H. and {Di Stefano}, R. 2003, ApJ, 590, L13

\bibitem[Kylafis \& Xilouris (1993)]{1993A&A...278L..43K} Kylafis, N. D. and Xilouris, E. M. 1993, A\&A, 278, 43


\bibitem[Liu et al. (2013)]{2013Natur.503..500L} Liu, J.-F., Bregman, J.N., Bai, Y., Justham, S., \& Crowther, P. 2013, Nature, 503, 500


\bibitem[Luangtip et al. (2016)]{2016MNRAS.460.4417L} Luangtip, W., Roberts, T. P., Done, C., 2016, MNRAS, 460, 4417

\bibitem[Lumb et al. (2002)]{2002A&A...389...93L} {Lumb}, D.~H., {Warwick}, R.~S., {Page}, M. and {De Luca}, A. 2002, A\&A, 389, 93

\bibitem[Marshall et al. (2002)]{2002ApJ...564..941M} {Marshall}, H.~L., {Canizares}, C.~R. and {Schulz}, N.~S. 2002, ApJ, 564, 941

\bibitem[Mezcua et al. (2016)]{2016ApJ...817...20M} Mezcua, M., Civano, F., Fabbiano, G., Miyaji, T., Marchesi, S. 2016, ApJ, 817, 20

\bibitem[Middleton et al. (2015b)]{2015MNRAS.454.3134M} Middleton, M. J., Walton, D. J., Fabian, A., Roberts, T. P., Heil, L., Pinto, C., et al. 2015, MNRAS, 454, 3134

\bibitem[Middleton et al. (2015a)]{2015MNRAS.447.3243M} Middleton, M. J., Heil, L., Pintore, F., Walton, D. J., and Roberts, T. P. 2015, MNRAS, 447, 3243

\bibitem[Middleton et al. (2014)]{2014MNRAS.438L..51M} Middleton, M. J., Walton, D. J., Roberts, T. P., Heil, L. 2014, MNRAS, 438, 51

\bibitem[Middleton et al. (2013)]{2013Natur.493..187M} Middleton, M.J., Miller-Jones, J.C.A., et al. 2013, Nature, 493, 187

\bibitem[Middleton et al. (2011)]{2011MNRAS.411..644M} Middleton, M.J., Roberts, T. P., Done, C., {Jackson}, F.~E. 2011, MNRAS, 411, 644

\bibitem[Miller et al. (2014)]{2014ApJ...788...53M} {Miller}, J.~M., {Raymond}, J., {Kallman}, T.~R., {Maitra}, D., {Fabian}, A.~C. et al. 2014, ApJ, 788, 53



\bibitem[Nandra et al. (2013)]{2013arXiv1306.2307N} Nandra, K., Barret, D., Barcons, X., Fabian, A. et al. 2013, The Hot and Energetic Universe, ArXiv=1306.2307

\bibitem[Pinto et al. (2016)]{2016Natur.533...64P} Pinto, C., Middleton, M.~J. and Fabian, A.~C. 2016, Nature, 533, 64 (Paper\,I)

\bibitem[Pinto et al. (2014)]{2014A&A...563A.115P} Pinto, C., Costantini, E., Fabian, A.~C., Kaastra, J. S. and in't Zand, J.~J.~M. 2014, A\&A, 563, 115

\bibitem[Pinto et al. (2013)]{2013A&A...551A..25P} Pinto, C., Kaastra, J. S., Costantini, E., de Vries, C. 2013, A\&A, 551, 25

\bibitem[Pinto et al. (2012)]{2012A&A...543A.134P} Pinto, C., {Ness}, J.-U., Verbunt, F., Kaastra, J. S., Costantini, E. and Detmers, R. 2012, A\&A, 543, 134

\bibitem[Pintore et al. (2015)]{2015MNRAS.448.1153P} {Pintore}, F., {Esposito}, P., {Zampieri}, L., {Motta}, S. and {Wolter}, A. 2015, MNRAS, 448, 1153

\bibitem[Porquet and Dubau (2000)]{2000A&AS..143..495P} {Porquet}, D. and {Dubau}, J. 2000 A\&AS, 143, 495

\bibitem[Poutanen et al. (2007)]{2007MNRAS.377.1187P} Poutanen, J., Lipunova, G., Fabrika, S., Butkevich, A. G., Abolmasov, P. 2007, MNRAS, 377, 1187

\bibitem[Roberts (2007)]{2007Ap&SS.311..203R} Roberts, T.P. 2007, Ap\&SS, 311, 203

\bibitem[Roberts et al. (2005)]{2005MNRAS.357.1363R} Roberts, T.P., Warwick R.S., Ward M.J., Goad M.R., \& Jenkins L.P. 2005, MNRAS, 357, 1363

\bibitem[Shakura and Sunyaev (1973)]{1973A&A....24..337S} Shakura, N. I. and Sunyaev, R. A. M. 1973, A\&A, 24, 337

\bibitem[Soria \& Kong (2016)]{2016MNRAS.456.1837S} Soria, R. and Kong, A. K. H. 2016, MNRAS, 456, 1837

\bibitem[Steenbrugge et al. (2003)]{2003A&A...402..477S} Steenbrugge, K. C., Kaastra, J. S., de Vries, C. P., \& Edelson, R. 2003, A\&A, 402, 477

\bibitem[Stobbart et al. (2004)]{2004MNRAS.351.1063S} Stobbart, A.-M., Roberts, T.P., \& Warwick, R. S 2004, MNRAS, 351, 1063

\bibitem[Stobbart et al. (2006)]{2006MNRAS.368..397S} Stobbart, A.-M., Roberts, T.P., \& Wilms, J. 2006, MNRAS, 368, 397

\bibitem[Sutton et al. (2013)]{2013MNRAS.435.1758S} Sutton, A. D., Roberts, T. P., Middleton, M. J. 2013, MNRAS, 435, 1758

\bibitem[Sutton et al. (2015)]{2015ApJ...814...73S} Sutton, A. D., Roberts, T. P., Middleton, M. J. 2015, MNRAS, 814, 73


\bibitem[Takeuchi et al. (2013]{2013PASJ...65...88T}Takeuchi, S., Ohsuga, K., Mineshige, S. 2013, PASJ, 65, 88

\bibitem[Turner et al. (2001)]{2001A&A...365L..27T} {Turner}, M.~J.~L., {Abbey}, A., {Arnaud}, M. et al. 2001, A\&A, 365, 27

\bibitem[van der Klis (1989)]{1989ARA&A..27..517V} van der Klis, M. 1989, ARA\&A, 27, 517

\bibitem[Volonteri et al. (2013)]{2013ApJ...775...94V} Volonteri, M., Sikora, M., Lasota, J.-P., Merloni, A. 2013, ApJ, 775, 94


\bibitem[Urquhart and Soria (2016)]{2016MNRAS.456.1859U} Urquhart, R. and Soria, R. 2016, MNRAS, 456, 1859

\bibitem[Uttley et al. (2014)]{2014A&ARv..22...72U} Uttley, P., Cackett, E. M., Fabian, A. C., Kara, E., Wilkins, D. R. 2014, A\&ARv, 22, 72

\bibitem[Uttley et al. (2011)]{2011MNRAS.414L..60U} Uttley, P. Cassatella, P., Wilkinson, T., Wilms, J., Pottschmidt, K. et al. 2011, MNRAS, 414, 60

\bibitem[Uttley \& McHardy (2001)]{2001MNRAS.323L..26U} Uttley, P. and McHardy, I. M. 2001, MNRAS, 323, 26

\bibitem[Vaughan et al. (2003)]{2003MNRAS.339.1237V} Uttley, P., Fabian, A. C. and Nandra, K. 2003, MNRAS, 339, 123

\bibitem[Walton et al. (2016a)]{2016ApJ...826L..26W} Walton, D.J., Middleton, M.J., Pinto, C., Fabian, A.C., Bachetti, M. et al. 2016, ApJ, 826, 26

\bibitem[Walton et al. (2016b)]{2016ApJ...827L..13W} Walton, D.J., Fuerst, F., Bachetti, M., {Barret}, D., 
	{Brightman}, M., {Fabian}, A.~C., et al. 2016, ApJ, 827, 13

\bibitem[Walton et al. (2016c)]{2016arXiv161006611W} Walton, D.J., Fuerst, F., Harrison, F.A., {Middleton}, M.~J., 
	{Fabian}, A.~C., {Bachetti}, M. et al. 2016, submitted to ApJ


\bibitem[Webb et al. (2012)]{2012Sci...337..554W} {Webb}, N., {Cseh}, D., {Lenc}, E., {Godet}, O., 
        {Barret}, D., {Corbel}, S., {Farrell}, S. et al. 2012, Science, 337, 554

\bibitem[Wilkinson \& Uttley (2009)]{2009MNRAS.397..666W} Wilkinson, T. \& Uttley, P. 2009, MNRAS, 397, 666

\end{thebibliography}

\bsp

\appendix

\section{Systematic effects}
\label{sec:appendix}

In this section we briefly discuss some systematic effects that might have 
limited our analysis, such as the contamination from the background
(instrumental and/or astrophysical) as well as the approximation 
of the ionization balance.
{We also provide some more technical detail on the selection
of flux intervals used to extract flux-resolved spectra 
and the plots of the power density spectra.
All this was shifted here to speed up the paper reading.}

\subsection{RGS background}
\label{sec:rgs_background}

\begin{figure}
  \includegraphics[width=0.95\columnwidth]{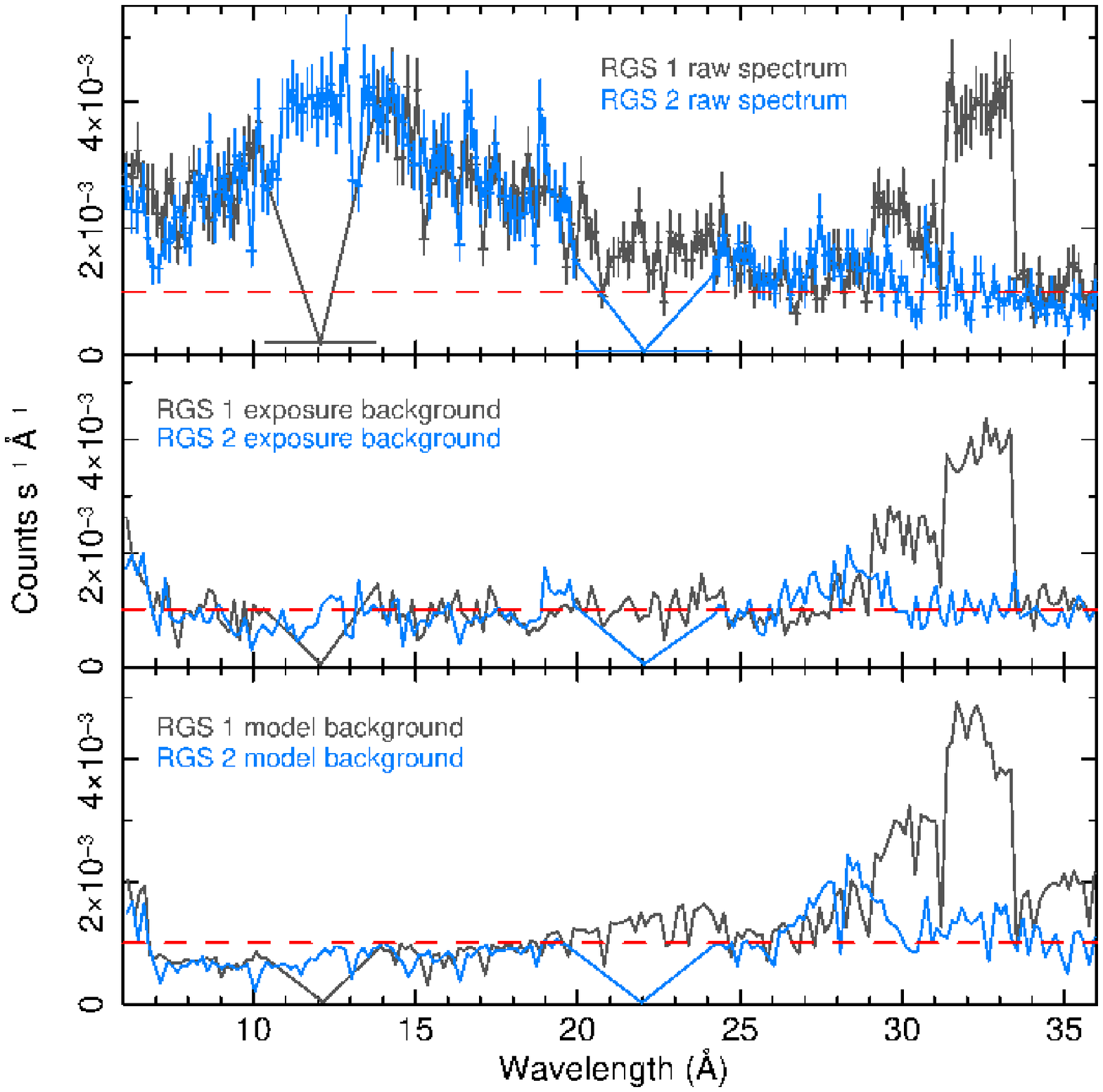}
   \caption{NGC 55 ULX XMM RGS 1-2 source raw spectra (top) with two alternative 
         background spectra: exposure-extracted background (middle)
         and blank-field-model background spectra (bottom).
         The background is severe below $\sim$\,7\,{\AA} and above $\sim$\,26\,{\AA}
         and higher for RGS\,1. RGS\,1 and RGS\,2 miss two chips covering
         the $10.4-13.8$\,{\AA} and $20-24$\,{\AA} range, respectively.} 
            \label{Fig:rgs_background}
\end{figure}

ULXs are faint sources if compared to common nearby Galactic X-ray binaries 
or active galactic nuclei. The EPIC-pn camera has a very large effective area,
which provides a lot of counts and an energy band broad enough to constrain the 
spectral continuum of ULXs. RGS spectrometers distribute the photons in much more
energy bins and therefore are much less sensitive. Searching for weak features 
in their spectra must therefore be a careful process, particularly 
when accounting for the background. 
Luckily, we have two ways of extracting an RGS background spectrum.
The first method uses the photons outside a certain cross-dispersion slit,
which according to the standard XMM-\textit{Newton}/SAS routine is the 98\%
of the PSF (see Fig.\,\ref{Fig:rgs_background}, middle panel).
Another method uses a model background estimated with ultradeep exposures 
of blank fields and scaled by the count rate in the RGS CCD 9 
where hardly any emission from the source is expected
(see Fig.\,\ref{Fig:rgs_background}, bottom panel).
The two background spectra are very similar and significantly dominate 
below $\sim$\,7\,{\AA} and above $\sim$\,26\,{\AA}. 
The RGS\,1 and 2 raw source spectra (and the spectral features) 
are significantly stronger than the background in our wavelength range.
For the work presented in this paper we have used the background
determined from the exposure, but we have repeated all the work 
using the model background and found no significant change in our results.
All this shows that the background has minor systematic effects
in our analysis.

\subsection{SED and ionization balance}
\label{sec:systematics_ionbal}

Another source of systematic uncertainties is the choice of the spectral
energy distribution (SED) since it may affect the ionization balance 
and therefore the calculation of ionization parameters 
and column densities in the photoionization components
(both in emission and absorption).

For instance, the standard $xabs$ model in SPEX adopts a Seyfert 1 type
of SED and ionization balance, whose shape of course deviates from the 
soft SED of a ULX such as NGC 55 ULX. 
To understand the effects of the SED choice, we test the photoionization $pion$ 
model in SPEX, which assumes the SED measured with the fitted spectral continuum model 
- the observed spectral shape of NGC 55 ULX - and instantaneously calculates
the photoionization equilibrium using the plasma routines available in SPEX 
(see the SPEX manual for more detail). 
At the moment the $pion$ emission model calculates the thermal emission of the photoionized layer 
by ignoring the photon-induced processes. This is not realistic enough, 
so for the moment we only check the effects of the SED on the absorption component. 

We have fitted again the RGS spectrum of NGC 55 ULX with the EPIC spectral continuum 
on top of which we have added a $pion$ model to substitute the $xabs$
component of the $0.2c$ absorber detected with highest significance. 
The main result is shown in Table\,\ref{table:xabs_pion_comparison}
along with the improvement in the fits previously obtained with 
the $xabs$ model (see also Table\,\ref{table:improvements}).
The results with the $xabs$ and the $pion$ models are consistent within the 
statistical uncertainties mainly due to the limited statistics.
However, there is no indication of major systematics
apart from the fact that the $pion$ model has larger uncertainties
due to the larger dependence of the parameters on the continuum shape.
The $pion$ model certainly provides a more physical description
of the absorber(s), but it requires much more memory and CPU time 
than the $xabs$ model, of course. The search for a 
multiple/continuous photoionization structure of the wind would
have been prohibitive with as many $pion$ components as the 
$xabs$ components used in Sect.\,\ref{sec:complex_wind_model} 
and Fig.\,\ref{Fig:wind_structure_IDL}.

\begin{table}
\caption{Systematic effects on the ionization balance.}  
\label{table:xabs_pion_comparison}             
\renewcommand{\arraystretch}{1.1}      
 \small\addtolength{\tabcolsep}{-0pt}
 
\scalebox{1}{%
\begin{tabular}{c|c c|c c c c c c c}     
\hline
Parameter                          & $xabs$ model            & $pion$ model        \\
\hline                                                                  
$N_{\rm H}$                        & $0.10\pm_{0.03}^{0.18}$ & $0.21\pm_{0.12}^{0.43}$  \\
$\log \xi$                         & $3.35 \pm 0.20$         & $3.44 \pm 0.29$  \\
$v_{\rm outflow}$                  & $0.199\pm0.003$         & $0.198\pm0.003$  \\
$\Delta\,\chi^2$/$\Delta$\,$C/dof$ & 19/18/3                 & 19/18/3          \\
\hline
\end{tabular}}

{ RGS fits improvements and details with of two alternative photoionized absorbers
  that use a default ($xabs$) and a measured ($pion$) ionization balance.
  The column densities are in standard units of $10^{24}$\,cm$^{-2}$.
  The photo-ionization parameters are in log ($\xi$, erg cm s$^{-1}$). 
  Outflow velocities are in units of light speed $c$.}

 \end{table}
         
\subsection{EPIC lightcurves for flux resolved spectra}
\label{sec:epic_flux_resolved_lighcurves}

In Sect.\,\ref{sec:epic_flux_resolved_spectra} we extracted EPIC spectra
in different flux ranges in order to understand the dependence 
of its spectral shape with the luminosity. 
Here we briefly describe the selection of the flux intervals.

\begin{figure*}
  \includegraphics[width=1.0\columnwidth]{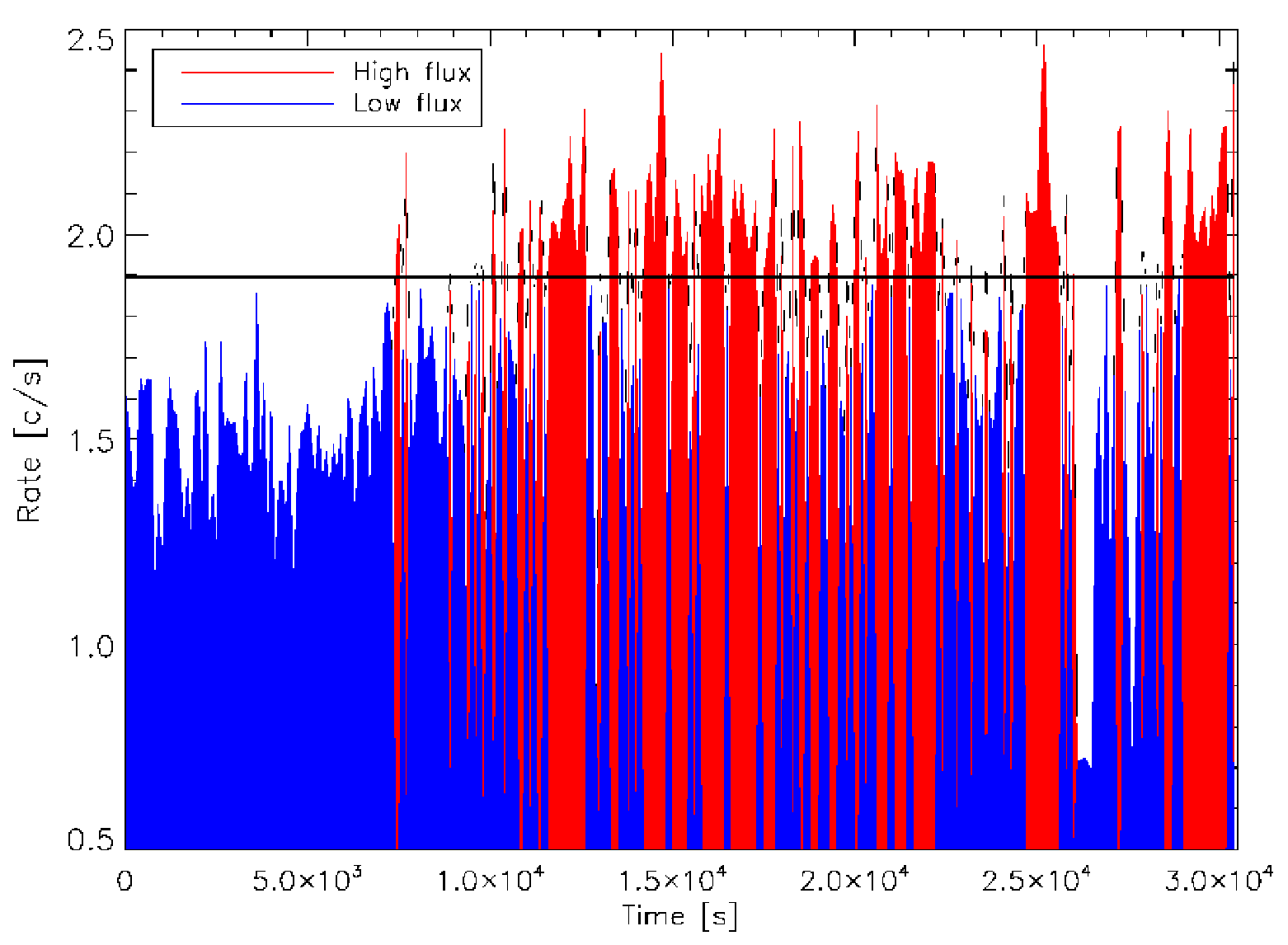}
  \includegraphics[width=1.0\columnwidth]{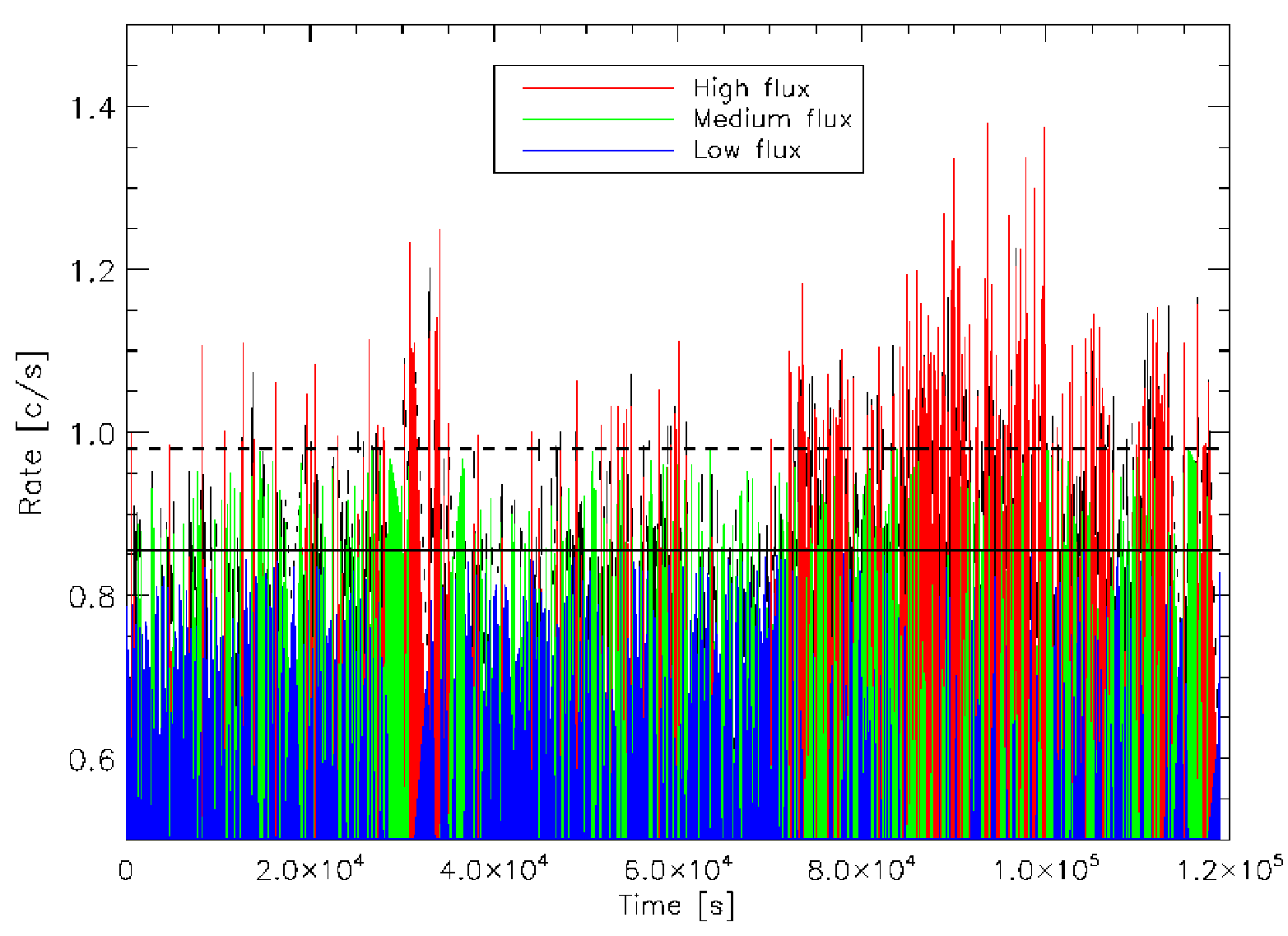}
   \caption{EPIC/pn 0.3-10 keV lightcurves of the first (left) 
            and second (right) observations
            both with time bins of 100\,s. 
            Flux thresholds were chosen in order have comparable statistics.
            For more detail see text.} 
            \label{Fig:lightcurves_flux_resolved}
\end{figure*}

At first, we extracted an EPIC/pn lightcurve in the 0.3--10\,keV energy band
for both the first (ID=0028740201) and the second exposure (ID=0655050101).
The source and background regions were the same as those used to extract
the spectra (see Sect.\,\ref{sec:data}). 
The background lightcurves were subtracted with the \textit{epiclccorr} task.
The background-subtracted lightcurves are shown in Fig.\,\ref{Fig:lightcurves_flux_resolved}.

We then split the two observation into five slices with comparable statistics.
In the first observation we selected bins with count rates above (or below) 1.895 count/s.
This provided two slices with 27 thousand counts each.
The second observation was instead split into three intervals of fluxes between
($<0.855$), (0.855--0.980), and ($>0.98$) counts/s.
This provided three slices with about 32 thousand counts each.
The statistics of these five time intervals are comparable. 
The five time intervals can be also seen in Fig.\,\ref{Fig:lightcurves_flux_resolved}.
We built the good time intervals using these intervals 
and extracted source and background spectra, response matrices,
and effective area auxiliary files with the \textit{evselect} task.
The flux-resolved spectra have been shown in Fig.\,\ref{Fig:epic_flux_resolved_spectra}.

\subsection{EPIC flux resolved spectra: Gaussian fits}
\label{sec:epic_gaussian_fits}

\begin{figure}
  \includegraphics[width=0.95\columnwidth]{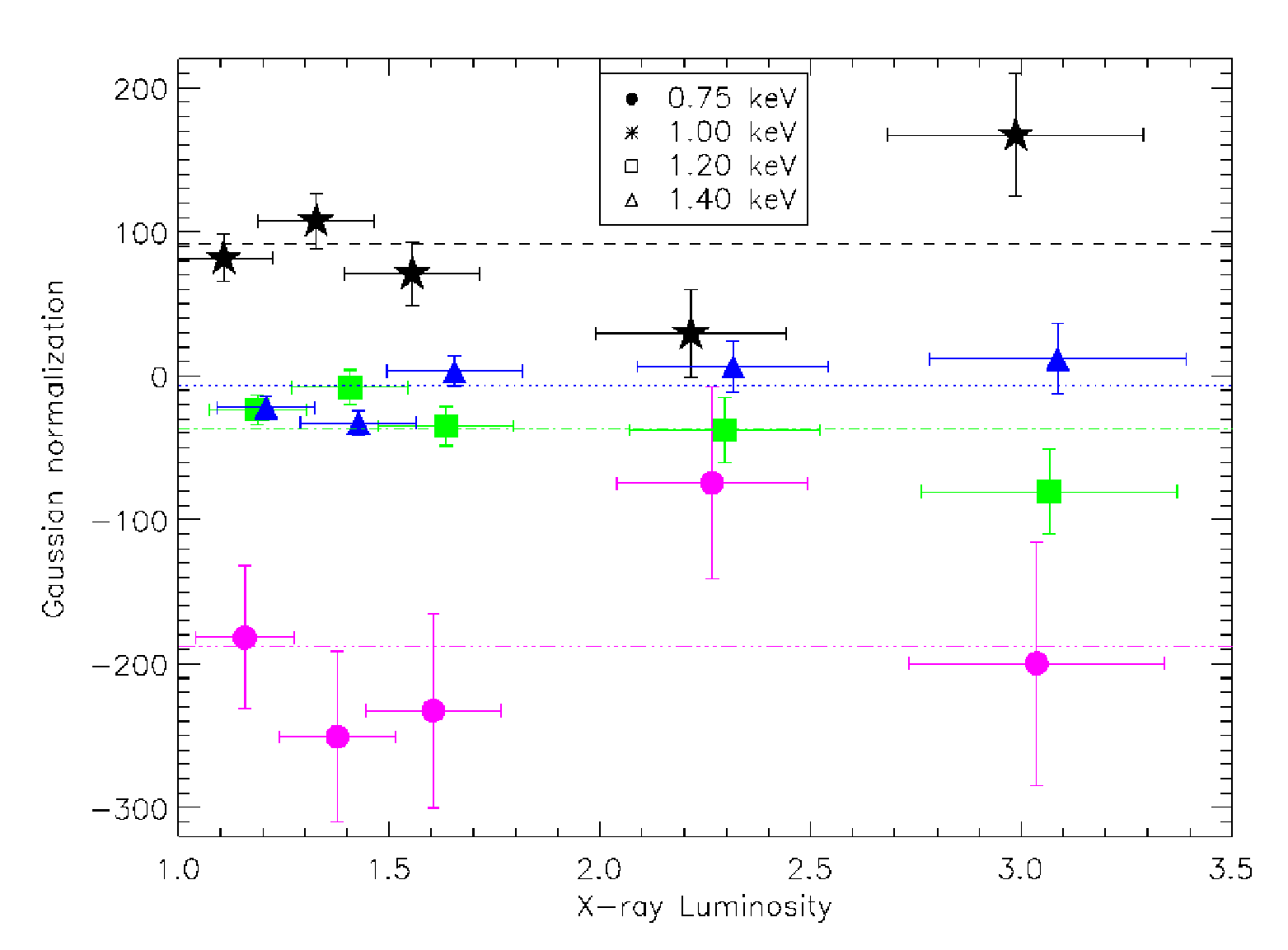}
   \caption{Gaussian fits to the flux-resolved EPIC pn spectra.
            Normalizations are in units of $10^{44}$ photons/s.
            Negative values refer to absorption lines.
            Luminosities are same as in Table\,\ref{table:epic_flux_resolved_spectra}.
            We slightly shifted the points along the X-axis for plotting purposes.} 
            \label{Fig:gaussians}
\end{figure}

We tested the behaviour of the EPIC spectral residuals by fitting a Gaussian line
for some of the strongest residuals (0.75, 1.0, 1.2, and 1.4\,keV, see Fig.\,\ref{Fig:epic_continuum}).
We adopted a FWHM of 100 eV similar to the EPIC spectral resolution.
We fit these four Gaussian lines with fix energies to the five flux-resolved spectra
and show the results in Fig.\,\ref{Fig:gaussians}.
The lines are broadly consistent within their mean values, but there are a few deviations.
Altough there is hint for variability, we notice that deeper observations are needed 
to confirm and study their variability in detail.

\subsection{Power density spectra}
\label{sec:power_spectra}

\begin{figure}
  \includegraphics[width=0.95\columnwidth]{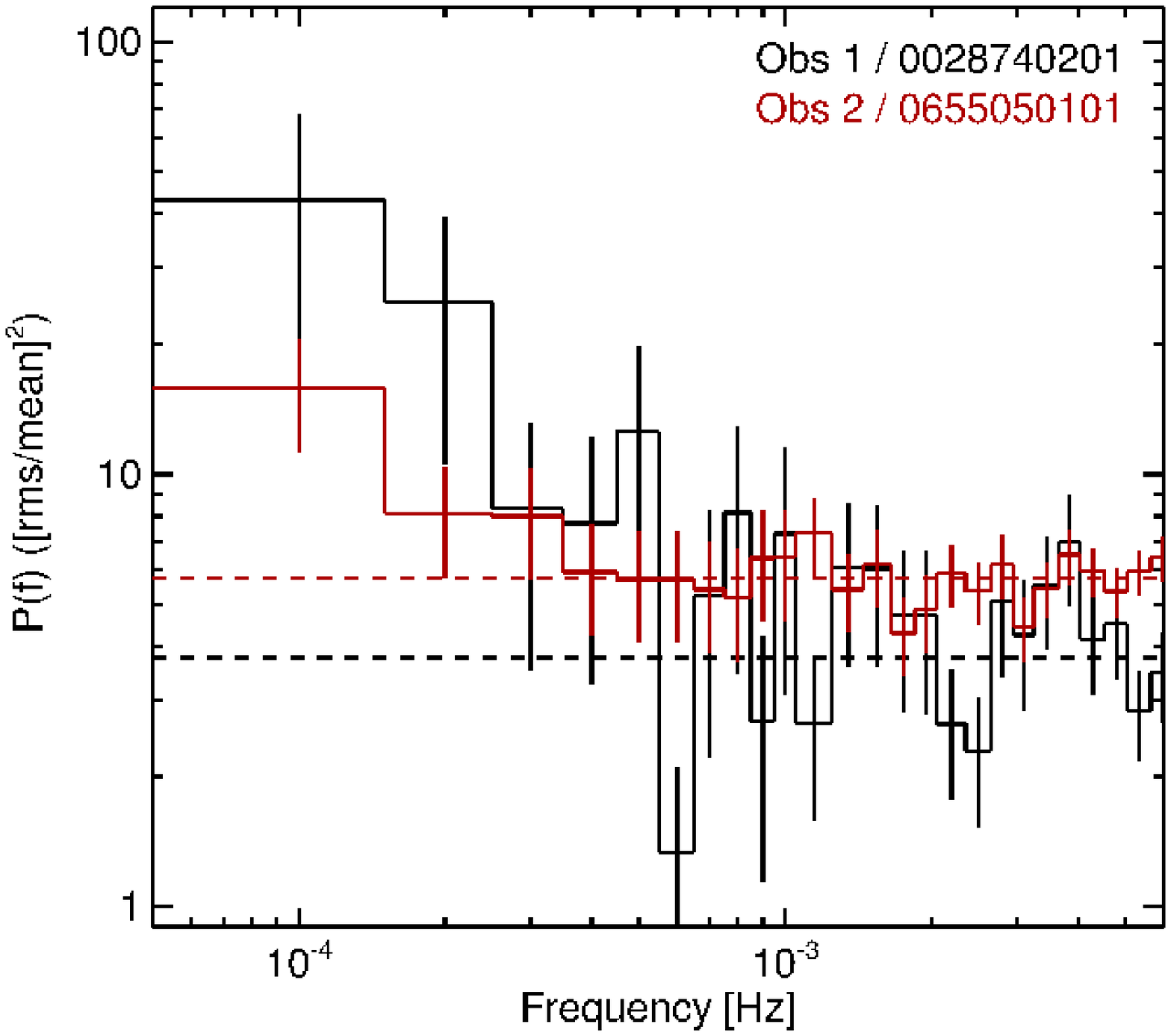}
   \caption{Power density spectra for the first (black) and 
   second (red) observations computed in the 0.3--1.0\,keV energy range. 
   The dashed lines show the Poisson noise level.} 
            \label{Fig:pds}
\end{figure}

We follow the standard method of computing a periodogram in 10 ks lightcurve 
segments before averaging over the segments at each frequency bin, 
and subsequent binning up over adjacent frequency bins 
(van der Klis 1989, Vaughan et al. 2003).
The resulting power density spectra are shown 
in Fig.\,\ref{Fig:pds}. Taking into account the level of Poisson noise,
given by the dashed lines, it shows that the source has very limited 
variability in 2010 compared to 2001. 

\subsection{Cross-correlation spectra}
\label{sec:ccf_spectra}

\begin{figure}
  \includegraphics[width=0.65\columnwidth, angle=90, bb=26 36 540 803]{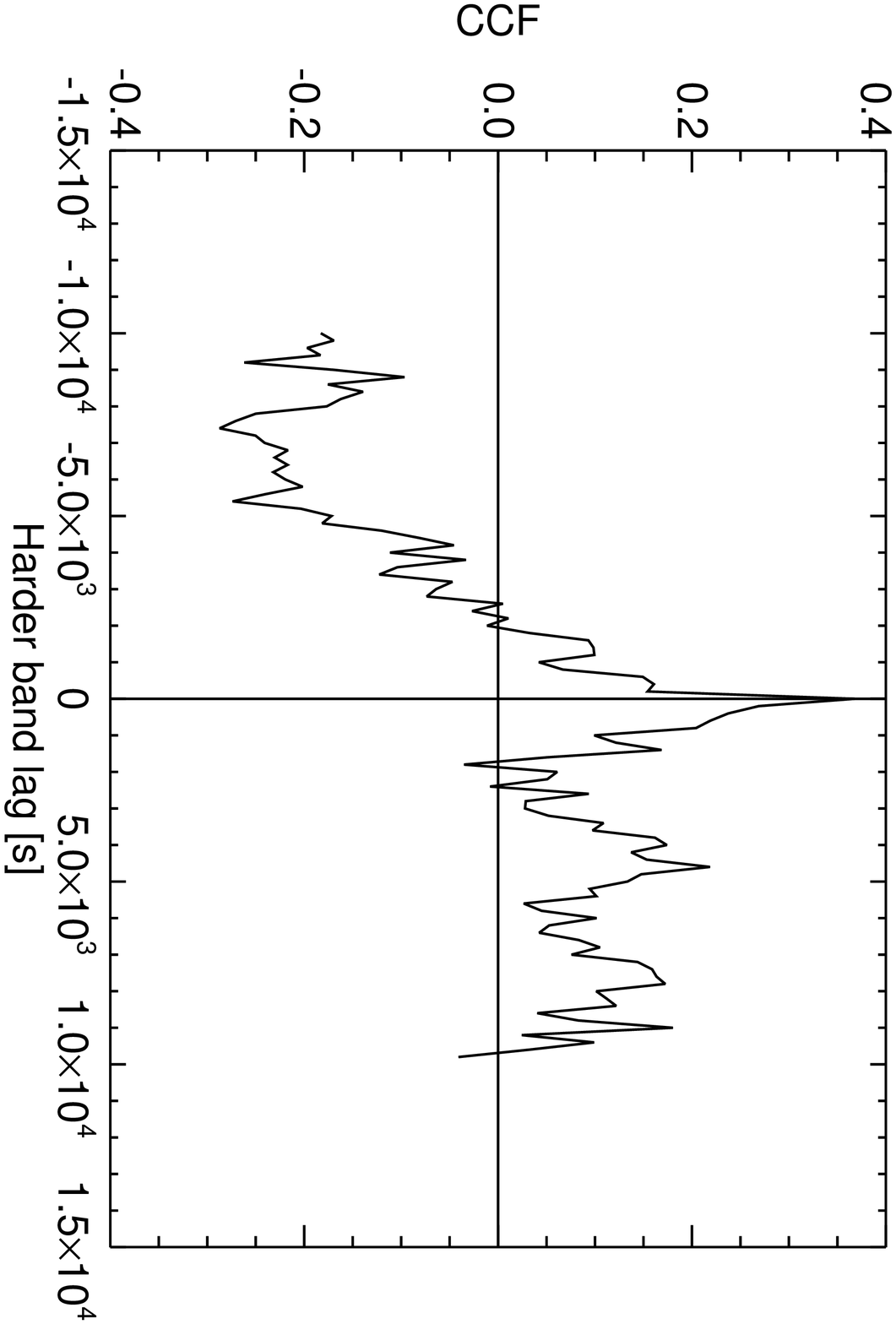}
  \includegraphics[width=0.65\columnwidth, angle=90, bb=26 36 540 803]{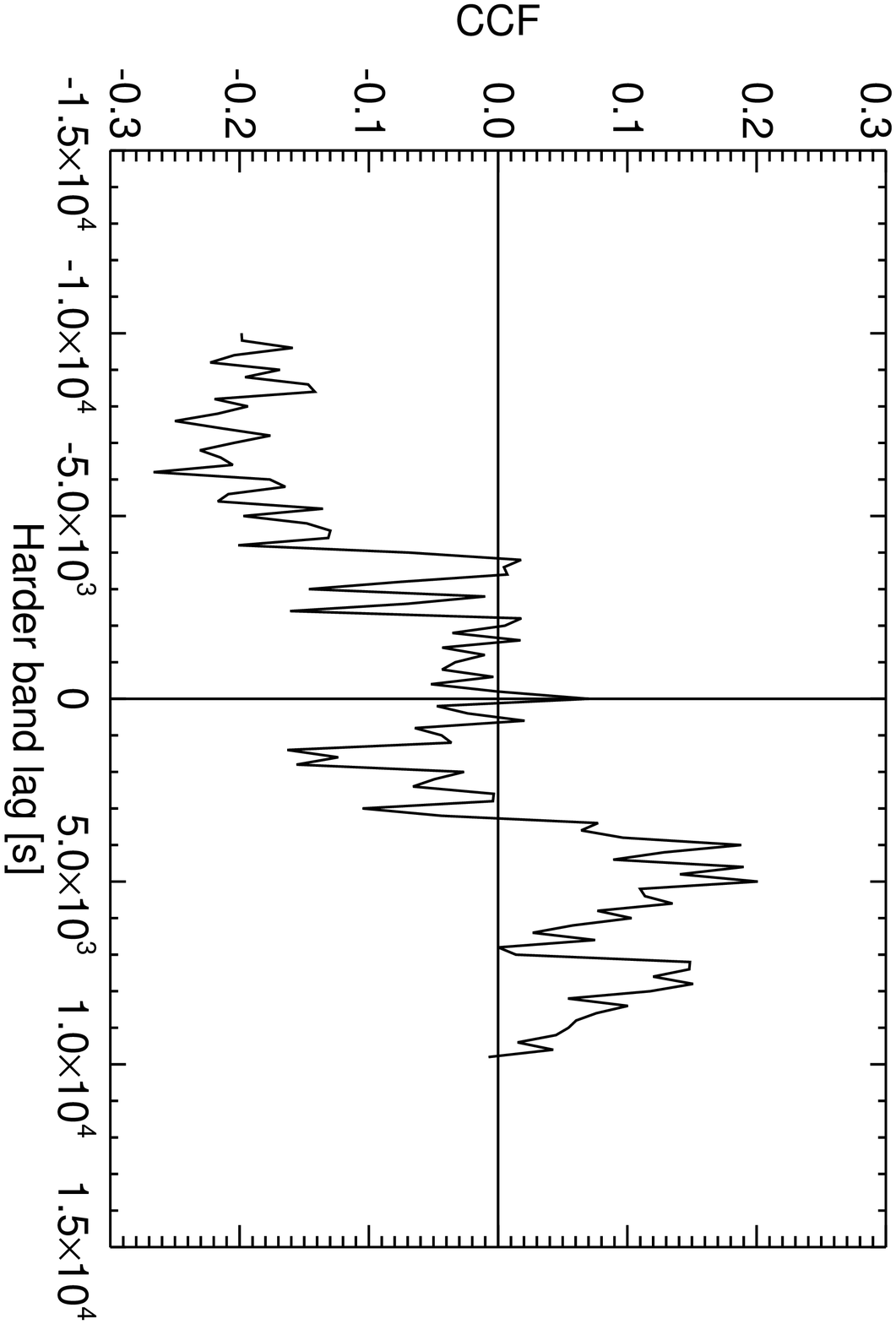}
   \caption{Cross-correlation (CCF) for the 
   second observation computed between the (0.3--1.0 vs 1.0--2.0\,keV, \textit{top}) and the 
   (0.3--1.0 vs 3.5--6.0\,keV, \textit{bottom}) energy bands. Notice the lag structure.} 
            \label{Fig:ccf_ob2_300_1000_1000_2000}
\end{figure}

For completeness, we have also computed the cross-correlation (CCF) between  
the soft energy band (0.3--1.0\,keV - same as reference 
band used in cross-spectrum, CS, throughout the paper) and several other bands. 
In Fig.\,\ref{Fig:ccf_ob2_300_1000_1000_2000} we show the CCF computed
with respect to the 1.0--2.0\,keV (top) and the 3.5--6.0\,keV (bottom) energy bands.
The harder band is always defined as having the positive lag on the x-axis.

The plot shows that there is clearly some lag structure in agreement with 
what we see in the CS. For the soft band separations, there is a strong correlation 
at zero time delay, with a skew towards a softer lag. 
As the energy band separation increases, the zero-lag correlation weakens, 
and we start to see a hard lag, as well as a negative lag. The changes seem to occur 
when looking at bands above $\sim$ 0.9-1.0 keV, similar to where we see the changes in the covariance.
The CCFs are however harder to interpret because they average over all the frequency dependent behavior.

\subsection{RGS spectra of other sources}
\label{sec:rgs_spectra_blazars}

In order to confirm that the lines detected in NGC 55 ULX are not produced by instrumental
features, we have searched for similar features in five extragalactic 
sources whose spectra are expected to be mainly featureless.
We chose these objects because the statistics quality of their RGS spectra is comparable 
to the RGS spectrum of NGC 55 ULX, once accounted for the net exposure time and the flux.
In Table\,\ref{table:noulx} we report the detail of the exposures used.
We repeat the RGS data reduction as done in Sect.\,\ref{sec:data} and the line search
as described in Sect.\,\ref{sec:line_detection} for these five sources.

\begin{figure*}
  \includegraphics[width=1.5\columnwidth,angle=90,bb=50 75 535 725]{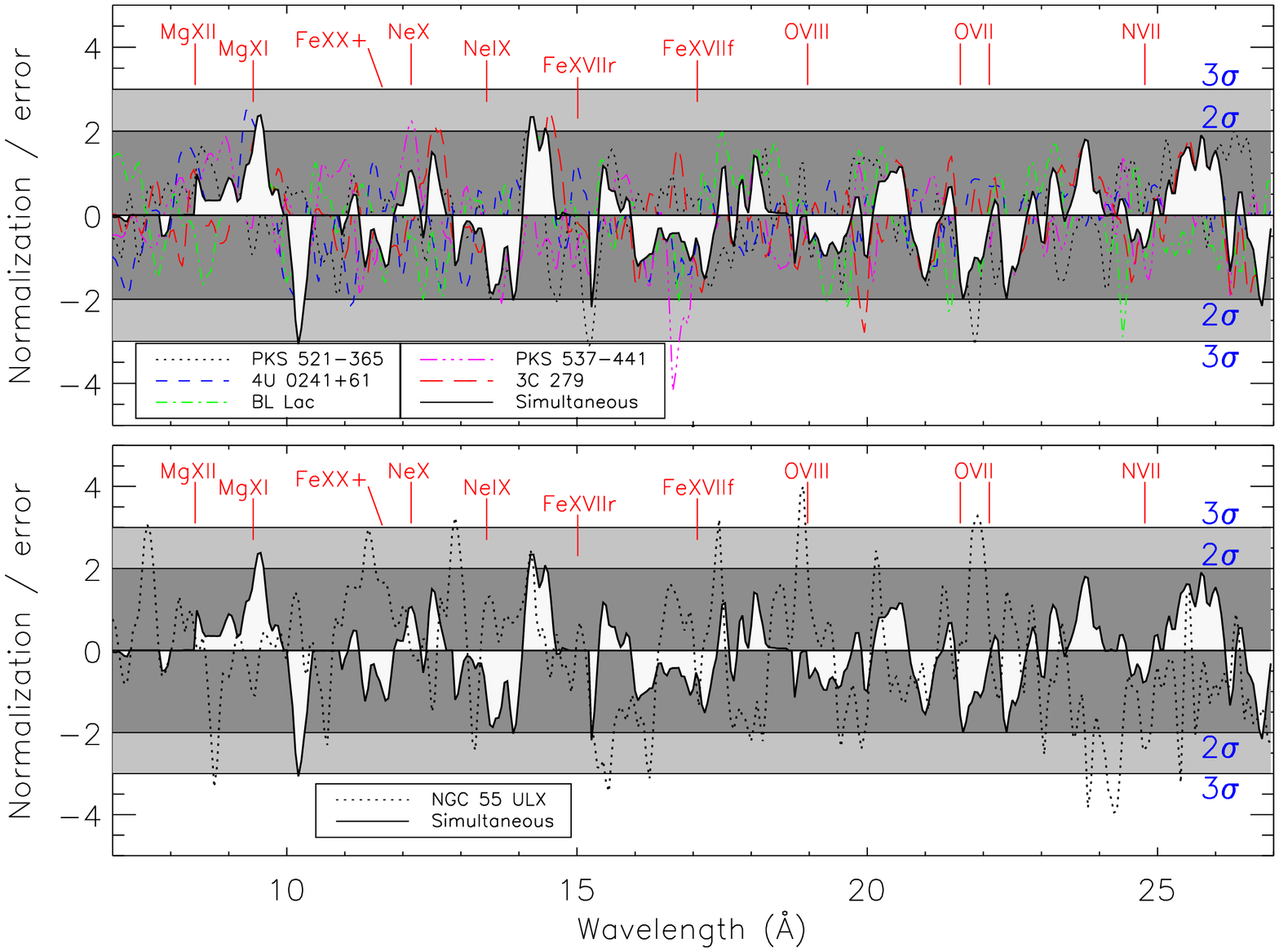}
   \caption{Line search for five power-law-like objects (top) and their simultaneous fits
   compared to that of NGC 55 ULX (bottom). The strong features detected in NGC 55 ULX 
   significantly differ from the weak instrumental features present in the other objects.} 
            \label{Fig:rgs_spectra_blazars}
\end{figure*}

In Fig.\,\ref{Fig:rgs_spectra_blazars} (top) we show the results obtained with the line-search
routine on each object adopting a power-law continuum corrected by redshift and Galactic absorption. 
In order to smear out any intrinsic features and strengthen the instrumental ones,
we have also simultaneously fitted the RGS spectra of the five objects, adopting the best-fit 
power-law continuum for each of them, whilst fitting the grid of Gaussian lines coupled 
between the models of the five spectra. In Fig.\,\ref{Fig:rgs_spectra_blazars} (bottom)
we compare the results from the line search obtained combining the five sources (solid line)
with that one obtained for NGC 55 ULX (dotted line).
At first, we notice that no emission feature is significantly detected at or above $3\sigma$,
but more importantly the strongest features detected in NGC 55 ULX are absent
in the other sources and in their combined analysis.
This shows that the lines detected in the RGS spectrum of NGC 55 ULX are significant 
and intrinsic to the source.

\begin{table}
\caption{XMM-\textit{Newton} observations of non-ULX sources.}  
\label{table:noulx}      
\renewcommand{\arraystretch}{1.1}
 \small\addtolength{\tabcolsep}{-1pt}
 
\scalebox{1}{%
\begin{tabular}{c c c c c c c c c c}     
\hline  
Source       &  Type     & ID          & t\,$_{\rm RGS1}^{(a)}$ & t\,$_{\rm RGS1}^{(b)}$   \\
             &           &             & (ks)                   & (ks)                     \\
\hline                                                                                   
PKS 521-365  & BL Lac    & 0065760201  &   31.9                 &    31.0                   \\
4U 241+61    & Seyfert 1 & 0503690101  &  118.4                 &    46.2                   \\
BL Lac       & BL Lac    & 0504370401  &  133.9                 &    83.5                   \\
PKS 537-441  & BL Lac    & 0551503301  &   24.5                 &    18.7                   \\
3C 279       & Quasar    & 0651610101  &  126.3                 &   107.7                   \\
\hline                
\end{tabular}}

$^{(a)}$ RGS nominal and $^{(b)}$ net exposure time after BKG screening. 
\end{table}

\label{lastpage}

\end{document}